\documentclass[a4paper,11pt]{article}

\usepackage{amsmath}
\usepackage{amssymb}
\usepackage{arrayjobx}
\usepackage[nosort]{cite}
\usepackage{subfigure}
\usepackage{graphicx}
\usepackage{color} 
\usepackage{bbm}
\usepackage[bulletsep]{collref}
\usepackage[margin=2cm]{caption}

\usepackage[latin1]{inputenc} 
\usepackage[T1]{fontenc}

\usepackage[bookmarks=true,hyperfigures=true]{hyperref}
 \usepackage[a4paper,text={470pt,700pt},centering]{geometry}
\makeatletter
\let\old@makecaption=\@makecaption
\def\@makecaption{\small\old@makecaption}
\makeatother
\providecommand{\hypersetup}[1]{}

\providecommand{\texorpdfstring}[2]{#1}
\providecommand{\pdfbookmark}[3][]{}

\hypersetup{plainpages=false}
\hypersetup{pdfpagemode=UseOutlines}
\hypersetup{bookmarksnumbered=true}
\hypersetup{bookmarksopen=true}
\hypersetup{pdfstartview=FitH}
\hypersetup{colorlinks=false}
 \hypersetup{citebordercolor={.5 1 .5}}
 \hypersetup{urlbordercolor={.5 1 1}}
 \hypersetup{linkbordercolor={1 .7 .7}}
\hypersetup{citebordercolor={.6 .9 .6}}
\hypersetup{urlbordercolor={.7 .8 1}}
\hypersetup{linkbordercolor={1 .7 .7}}
\hypersetup{pdfborder={0 0 1 [3]}}

\numberwithin{equation}{section}
\newcommand{\vev}[1]{\langle #1\rangle}

\newcommand{\jjs}{c^s_1}
\newcommand{\jsj}{c^s_2}

\newcommand{\jsjsab}{c^{ss}_1}
\newcommand{\jsjsba}{c^{ss}_2}

\newcommand{\jjb}{d_1}
\newcommand{\jbj}{d_2}

\newcommand{\jsjsb}{d_1^s}
\newcommand{\jsbjs}{d_2^s}

\ifx\genfrac\sdflkaj

\else

\fi

\newcommand{\p}{\partial}
\newcommand{\pb}{\bar \partial}

\newcommand{\alg}[1]{\mathfrak{#1}}

\newcommand{\nn}{\nonumber}

\makeatletter
\def\mr@ignsp#1 {\ifx\:#1\@empty\else #1\expandafter\mr@ignsp\fi}%
\newcommand{\multiref}[1]{\begingroup
\xdef\mr@no@sparg{\expandafter\mr@ignsp#1 \: }%
\def\mr@comma{}%
\@for\mr@refs:=\mr@no@sparg\do{\mr@comma\def\mr@comma{,}\ref{\mr@refs}}%
\endgroup}
\makeatother

\newcommand{\hypref}[2]{\ifx\href\asklfhas #2\else\href{#1}{#2}\fi}

\newcommand{\secref}[1]{Sec.~\multiref{#1}}

\newcommand{\figref}[1]{Fig.~\multiref{#1}}
\renewcommand{\eqref}[1]{(\multiref{#1})}

\newcommand{\hJ}{\mathcal{J}}
\newcommand{\ahJ}{\bar{\mathcal{J}}}

\def\[{\begin{equation}}
\def\]{\end{equation}}
\def\<{\begin{eqnarray}}
\def\>{\end{eqnarray}}
\newcommand{\beq}{\begin{equation}}
\newcommand{\eeq}{\end{equation}}
\newcommand{\beqa}{\begin{eqnarray}}
\newcommand{\eeqa}{\end{eqnarray}}

\begin{document}

\thispagestyle{empty}


\begin{flushright}
{\tt HU-MATH-2016-18}\\
{\tt TCDMATH 16-14}
\end{flushright}

\begin{center}
{\Large \bfseries 
Multi-Soft gluon limits and extended current algebras\\
at null-infinity
\par}%
\vspace{10mm}

\begingroup\scshape\large 
Tristan McLoughlin${}^{1}$ and Dhritiman Nandan${}^{2,3,4}$
\endgroup
\vspace{5mm}

\textit{${}^{1}$ School of Mathematics, Trinity College Dublin\\ College Green, Dublin 2, Ireland}\\[0.1cm]
  \texttt{\small tristan@maths.tcd.ie\phantom{\ldots}} \\ \vspace{4mm}
  
\textit{${}^{2}$ Institut f\"ur Physik und IRIS Adlershof, Humboldt-Universit\"at zu Berlin \phantom{$^\S$}\\
  Zum Gro{\ss}en Windkanal 6, 12489 Berlin, Germany} \\[0.1cm]
\texttt{\small dhritiman@physik.hu-berlin.de\phantom{\ldots}} \vspace{4mm}

\textit{${}^{3}$ Institut f\"ur Mathematik und IRIS Adlershof, Humboldt-Universit\"at zu Berlin \phantom{$^\S$}\\
  Zum Gro{\ss}en Windkanal 6, 12489 Berlin, Germany} \\[0.1cm]
  \vspace{4mm}

\textit{${}^{4}$ Higgs Centre for Theoretical Physics, School of Physics and Astronomy,\phantom{$^\S$}\\
	The University of Edinburgh, Edinburgh EH9 3JZ, Scotland, UK} \\[0.1cm]
\texttt{\small dhritiman.nandan@ed.ac.uk\phantom{\ldots}}
\vspace{4mm}

\textbf{Abstract}\vspace{5mm}\par
\begin{minipage}{14.7cm}

In this note we consider aspects of the current algebra interpretation of multi-soft limits of tree-level gluon scattering amplitudes
in four dimensions. Building on the relation between a positive helicity gluon soft-limit and the Ward identity for a level-zero Kac-Moody
current, we use the double-soft limit to define the Sugawara energy-momentum tensor and, by using the triple- and quadruple-soft limits, 
show that it satisfies the correct OPEs for a CFT. We study the resulting Knizhnik-Zamolodchikov equations and show that they hold for positive helicity gluons
in MHV amplitudes. Turning to the sub-leading soft-terms we define a one-parameter family of currents whose Ward identities correspond to the universal
tree-level sub-leading soft-behaviour. We compute the algebra of these currents formed with the leading currents and amongst themselves. Finally, 
by parameterising the ambiguity in the double-soft limit for mixed helicities, we introduce a non-trivial OPE between the holomorphic and
anti-holomorphic currents and study some of its implications.

\end{minipage}\par
\end{center}
\newpage


\setcounter{tocdepth}{4}
\hrule height 0.75pt
\tableofcontents
\vspace{0.8cm}
\hrule height 0.75pt
\vspace{1cm}

\setcounter{tocdepth}{2}


\section{Introduction}

The simple geometric fact that a null vector in Minkowski space defines a point on the
sphere at null infinity naturally suggests interpreting scattering amplitudes for massless particles 
as two-dimensional correlation functions. An early and  prominent example of such an interpretation was given
by Nair in \cite{Nair:1988bq} where certain ${\cal N}=4$ super-Yang-Mills (SYM) amplitudes were constructed
using the operator product expansion (OPE) of the Kac-Moody currents which define a two-dimensional Wess-Zumino-Witten model. 
This construction in part lead to the development of the twistor string \cite{Witten:2003nn}
which gives a complete description of tree-level ${\cal N}=4$ SYM in terms of a two-dimensional
world-sheet theory.

Kac-Moody structures have also recently appeared in the study of the soft-limits of gluon amplitudes \cite{Strominger:2013lka, He:2015zea}. 
Scattering amplitudes of massless particles such as gluons can be formally expanded in powers of a small scaling parameter
$\delta$ multiplying the momentum of a particle which is thus taken to be soft, $p^{\mu}=\delta\, q^{\mu}$, and which schematically gives 
\<
\lim_{\delta\to 0} {\cal A}_{n+1}(\delta q) = \Bigl (\frac 1 \delta \, S^{(0)}(q)+ S^{(1)}(q)
\Bigr)\, {\cal A}_{n}
+ {\cal O}(\delta) \, ,
\>
where ${\cal A}_{n}$ depends only on the remaining hard momenta. 
That scattering amplitudes of photons and gravitons have a universal, leading divergent behaviour
as the external leg becomes soft has been long understood
\cite{Low:1954kd, GellMann:1954kc, Low:1958sn, Weinberg:1964ew}.  It is also known that at tree-level the
behaviour at the sub-leading order is universal, which is sometimes referred to as the Low theorem
 \cite{Low:1954kd, GellMann:1954kc, Low:1958sn, Burnett:1967km,  Gross:1968in, Jackiw:1968zza}, while for tree-level 
 graviton amplitudes there is further universal behaviour at sub-sub-leading order \cite{Cachazo:2014fwa}. 
 Analogous sub-leading behaviour for gluon amplitudes has been 
 recently studied  \cite{Casali:2014xpa, Schwab:2014xua, Bern:2014vva, Broedel:2014fsa, Larkoski:2014bxa, White:2014qia}. 
In \cite{He:2015zea}  He \textit{et al}  showed 
that one could identify the tree-level leading soft theorem for  positive helicity gluons with the Ward identity for 
holomorphic two-dimensional Kac-Moody currents.  These currents were shown to be related to
the asymptotic symmetries of Yang-Mills theory given by large CPT-invariant gauge
 transformations \cite{Strominger:2013lka}\footnote{We focus entirely on gluon amplitudes but the relation between asymptotic symmetries and soft-behaviour 
has been the subject of significant 
 recent interest particularly for gravitons and photons, see for example \cite{Strominger:2013jfa, He:2014laa, Kapec:2014opa, Campiglia:2014yka},
\cite{Kapec:2015vwa, Campiglia:2016efb, Afkhami-Jeddi:2014fia, Zlotnikov:2014sva, Kalousios:2014uva},
  \cite{He:2014cra, Lysov:2014csa, Kapec:2014zla, Kapec:2015ena, Conde:2016csj}.}. The OPE of the currents was extracted
 from the limit of two soft positive helicity gluons and it was shown that the level is zero in this identification, that is the algebra is a 
 standard current algebra
 \<
\label{eq:JJ}
J^a(z_1) J^b(z_2)\sim~ 
\frac{i f^{ab}{}_{c} J^c(z_2)}{z_{12}}~.
\>
 It was further noted that the limit where one positive and one negative helicity gluon are both taken soft
 is ambiguous with the result depending on the order in which the
 limit is taken. In  \cite{He:2015zea}  the
 authors gave the prescription of first taking all positive helicity gluons soft which 
 resulted in one copy of the holomorphic Kac-Moody algebra but without a second anti-holomorphic 
 copy arising from the negative helicity gluons. 
 
 Multi-soft limits of gluon amplitudes have been studied more generally in a number of recent works and
 universal behaviour in the limit where all gluons are taken soft
simultaneous was found for tree-level amplitudes using the BCFW \cite{Britto:2005fq} 
and CHY \cite{Cachazo:2013iea,Cachazo:2013hca} formalisms in  \cite{Klose:2015xoa, Volovich:2015yoa}. These limits have also been studied using the
CSW \cite{Cachazo:2004kj} formalism by \cite{Georgiou:2015jfa} and by considering the field theory limit of string theory \cite{DiVecchia:2015bfa}. At leading
order in the soft parameter this simultaneous double- and multi-soft limit differs from the consecutive case when there are 
particles of both positive and negative helicity. There is also universal behaviour in the sub-leading order of the  multi-soft limit
however the order in which the limit is taken is important even when all gluons have the same helicity.

We use these multi-soft limits to study further aspects of the current algebra interpretation 
 in four dimensions. Starting with the leading soft-limit for positive helicity gluons we
show that, by analogy with the Sugawara construction \cite{Sugawara:1967rw}, one can use the double soft-limit to define a
holomorphic energy-momentum tensor. Moreover we show, by analyzing the triple- and
quadruple-soft limits of positive helicity gluons that this energy-momentum tensor has the standard OPE
with both the holomorphic current and with itself. In a conformal field theory with a 
Kac-Moody symmetry the correlation functions of primary fields should satisfy certain
differential equations, the Knizhnik-Zamolodchikov (KZ) equations \cite{Knizhnik:1984nr}.
We consider the KZ equation for the amplitudes 
and show that while they hold for positive helicity gluons in MHV amplitudes
this is not the case for the negative helicity gluons or for non-MHV amplitudes. 
While this suggests that the holomorphic currents do not provide a complete description of the 
putative CFT it motivates further analysing the sub-leading and negative helicity soft-limit. 

We thus study the soft limit at sub-leading order and give these sub-leading soft-theorems an interpretation as
Ward identities for two-dimensional currents which depend on a continuous
parameter corresponding to the soft gluon energy, ${J}_{\rm sub}^a(z;\omega_{z})$.
 The OPE of these currents with themselves and with the leading order
currents is then extracted from the sub-leading double-soft limits. 
As the algebra depends on exactly how the double soft limit is 
taken we introduce parameters to characterise this ambiguity. In the language of the two-dimensional field theory the leading terms of the OPEs are 
given by
\<
\label{eq:JJsub}
{J}^a(z_1) {J}_{\rm sub}^b(z_2;\omega_{z_2})\kern-5pt  &\sim& \kern-5pt  i f^{ab}{}_{c}    \frac{J_{\rm sub}^c(z_2;  \omega_{z_2})}{z_{12}}~,
\>
\<
\label{eq:JsubJsub}
{J}_{\rm sub}^a(z_1;\omega_{z_1}){J}_{\rm sub}^b(z_2;\omega_{z_2}) \kern-5pt& \sim &\kern-5pt \frac{i f^{ab}{}_{c} J_{\rm sub}^c(z_2;\jsjsba   \omega_{z_1}+\jsjsab \omega_{z_2})}{z_{12}}~.
\>

As in \cite{He:2015zea} the anti-holomorphic currents are defined using the soft-limits of negative helicity gluons and to study the 
algebra with the holomorphic currents we consider double-soft limits involving both a positive and negative 
helicity gluon. However rather than picking a specific ordering of soft-limits we consider a linear combination and introduce variables with which we 
parameterise the ambiguity. The resulting algebra of currents is found to be
\<
\label{eq:JJbar}
{J}^a(z_1)\bar{{J}}^b(z_2)\kern-5pt & \sim&\kern-5pt if^{ab}{}_{c}\Big[\jjb \frac{\bar{J}^c(z_2)}{z_{12}}-\jbj \frac{{J}^c(z_2)}{\bar {z}_{1 2}}
-\jbj  \frac{z_{12}}{\bar {z}_{12}}\p J^c(z_2)\Big] ~.
\>
This OPE differs from the usual CFT chiral current algebra where one would find ${J}^a(z_1)\bar{{J}}^b(z_2)\sim 0$. However such OPEs can be considered in general \cite{Luscher:1977uq} and bear resemblance to the non-chiral 
current algebras that appear in the study of super-group CFTs
\cite{Ashok:2009xx}. Of course there are a number of differences, not least the absence of logarithmic terms, at least at tree-level, and naturally the Killing form for the gauge group in Yang-Mills theory is not usually taken to vanish as in the super-group case.  
To complete the algebra of currents we calculate the OPEs of the sub-leading currents with the anti-holomorphic currents and sub-leading anti-holomorphic coordinates
\<
\bar{{J}}^a(z_1) {J}_{\rm sub}^b(z_2;\omega_{z_2})\kern-5pt & \sim  &\kern-5pt if^{ab}{}_{c} 
 \frac{  \bar{J}_{\rm sub}^c(z_2; \omega_{z_2})}{\bar z_{12} }~,
\>
\<
{{J}}_{\rm sub}^a(z_1;\omega_{z_1}) \bar{J}_{\rm sub}^b(z_2;\omega_{z_2})\kern-5pt  &\sim&\kern-5pt
  i f^{ab}{}_{c}\Big[
\frac{ {J}_{\rm sub}^c(z_2; \jsbjs   \omega_{z_2})}{\bar z_{12} }
-\frac{ \bar{J}_{\rm sub}^c(z_2; \jsjsb  \omega_{z_1})}{z_{12} }\nn\\
 & & \kern+30pt +\frac{  \bar z_{12}}{z_{12} }\pb \bar{J}_{\rm sub}^c(z_2; \jsjsb   \omega_{z_1})
+2 \frac{  z_{12}}{\bar z_{12} } \p {J}_{\rm sub}^c(z_2; \jsbjs  \omega_{z_2}) \Big]~.
\>
Using \eqref{eq:JJbar} one can compute the OPE of the anti-holomorphic currents with the 
holomorphic stress-energy tensor. We do this in \secref{sec:JbCFT} and show that it is consistent with the triple-soft limit involving two positive helicity gluons and one negative. Finally, using the leading order holomorphic and anti-holomorphic currents, we define a conjugation operator $C(z)\propto\, :\!J^a\bar J^a\!:\!(z)$ 
and compute its OPE with the currents. 

While we don't attempt to give a more fundamental interpretation of these two-dimensional currents the renewed focus on understanding the asymptotic symmetries
has lead to a number of interesting proposals for a holographic description of flat space scattering amplitudes. In \cite{Adamo:2015fwa}, building on \cite{Adamo:2014yya}, a world-sheet theory with the null infinity boundary of flat space as its target space was proposed. An alternative approach based on the ambi-twistor string, \cite{Mason:2013sva,Geyer:2014fka}, was studied in \cite{Geyer:2014lca, Lipstein:2015rxa}. Recently, Cheung \textit{et al} \cite{Cheung:2016iub} have proposed a holographic description which following, \cite{deBoer:2003vf, Solodukhin:2004gs}, makes use of a slicing of flat space into a family of AdS$_3$ sub-spaces to interpret four-dimensional 
amplitudes in terms of a two-dimensional CFT. They identify the conserved currents and energy-momentum tensor with the asymptotic symmetries of 
gluon and graviton amplitudes. While there are a number of differences with our current work, for example the central charge of the current algebra, 
it would be very interesting to understand better if there is a holographic interpretation of the sub-leading and anti-holomorphic currents. 

\section{Soft limits for gluon amplitudes}
The central objects of our study are the soft limits of gluon scattering amplitudes. While
we will be interested in the full colour dressed amplitudes it is convenient to start with
colour stripped amplitudes. 
Consider an $n\!+\!1$-leg amplitude where we take the first  leg to be a soft particle  of helicity $h_{q}$ with 
momentum $\delta q$, or in terms of spinor variables, $\{\sqrt{\delta} \lambda_{q}, \sqrt{\delta} {\tilde \lambda}_{q}\}$
\footnote{We will make extensive use of the the spinor-helicity formalism. Useful reviews can be found in 
\cite{Dixon:1996wi, Henn:2014yza, Elvang:2013cua}.} .
The soft limit corresponds to expanding the
amplitude in powers of $\delta_1$ and we keep only those terms in the expansion which are universal. As it will make subsequent formulae simpler we 
will often simply denote
the non-soft legs with momenta $p_i$, $i=1,\dots, n$ by $1, 2, \dots, n $. 

\subsection{Single-soft limits}
\label{sec:ss}
The single-soft limits for gluon amplitudes were found, up to the first sub-leading term, in
\cite{Low:1958sn,Burnett:1967km,Casali:2014xpa}
\<
{ A}_{n+1}(\delta_1 q^{h_{q}}, 1, \dots, n)\rightarrow \left(\frac{1}{\delta_1} S^{(0)}_{n,1}( q^{h_{q}})+S^{(1)}_{n,1}( q^{h_{q}})\right){A}_n(1, \dots, n)
\>
where, for a positive helicity gluon, $ h_{q}=+ $, between neighboring particles $ n $ and $ 1 $, the universal soft-factors can be written using the spinor-helicity
formalism as 
\<
\label{eq:single_soft}
S^{(0)}_{n,1} (q^+)=\frac{\langle n 1\rangle }{\langle n~ q\rangle \langle q 1\rangle}~, ~~~S^{(1)}_{n,1}( q^+)=\frac{1}{\langle q 1\rangle}{\tilde \lambda}_{q}^{\dot \alpha} \frac{\partial}{\partial{\tilde \lambda}^{\dot \alpha}_{1}}+\frac{1}{\langle n~ q\rangle  }{\tilde \lambda}_{q}^{\dot \alpha} \frac{\partial}{\partial{\tilde \lambda}^{\dot \alpha}_{n}}~.
\>
 Here we use the conventions
\<
\langle{a b}\rangle=\epsilon_{\alpha \beta}\lambda_a^\alpha\lambda_b^\beta=\lambda_a{}_\beta\lambda_b^\beta
=-\lambda_a^\alpha\lambda_b{}_\alpha
\>
and similarly for the dotted indices, $[ab]=\epsilon_{\dot \alpha \dot \beta}\tilde \lambda_a^{\dot \alpha}\tilde \lambda_b^{\dot \beta}$. Correspondingly we will write
the contraction of a spinor with a derivative as 
\<
{\tilde \lambda}_{a}^{\dot \alpha} \frac{\partial}{\partial{\tilde \lambda}^{\dot \alpha}_{b}}=-[a\, \pb_b]~.
\>
 For a negative helicity gluon the soft factors are given by conjugation of the spinor variables, $\lambda_i \leftrightarrow {\tilde \lambda}_i$.


\subsection{Multi-soft limits}
One can similarly consider the limit of gluon amplitudes with multiple soft gluons. However as mentioned there is in general an ambiguity in the result 
which depends on the order in which the gluons are taken to be soft.  
Starting with the case of two soft gluons we consider the $n\!+\!2$-leg amplitude where we will take the first and second legs to be soft particles with helicities $h_{q_1}$ and 
$h_{q_2}$ and
momenta $\{\sqrt{\delta_1} \lambda_{q_1}, \sqrt{\delta_1} {\tilde \lambda}_{q_1}\}$ and $\{\sqrt{\delta_2} \lambda_{q_2}, \sqrt{\delta_2} {\tilde \lambda}_{q_2}\}$ 
respectively. In particular, if one takes the gluons soft sequentially 
we call this a consecutive soft limit in contradistinction to the simultaneous or double soft limit. 
These consecutive soft limits can be calculated straightforwardly from repeated
action of the above single soft factors. 

We define the consecutive soft factor, ${\tt CSL}_{n,1}(q_1^{h_{q_1}}, q_2^{h_{q_2}})$, to be
\<
{\tt CSL}_{n,1}(q_1^{h_{q_1}}, q_2^{h_{q_2}}){A}_n(1,\dots, n)&\equiv& \lim_{\delta_2 \to 0} \lim_{\delta_1 \to 0}{A}_{n+2}(\delta_1 q^{h_{q_1}}_1, \delta_2 q^{h_{q_2}}_2, 1 \dots, n) \nn\\
& &\kern-140pt =
\left(\frac{1}{\delta_1} S^{(0)}_{n,q_2}(q_1^{h_{q_1}})+S^{(1)}_{n,q_2}(q_1^{h_{q_1}})\right)
\left(\frac{1}{\delta_2} S^{(0)}_{n,1}(q_2^{h_{q_2}})+S^{(1)}_{{n,1}}(q_2^{h_{q_2}})\right) {A}_n(1,\dots, n)~,
\>
where we take the particle $q_1$ soft first and then $q_2$. 
As it will be of interest later, let us give the explicit expression:
\<
 {\tt CSL}_{n,1}(q_1^{+}, q_2^{+})&=&\frac{1}{\delta_1\delta_2}\frac{\langle n\, 1\rangle}{\langle n q_1\rangle  \langle q_1 q_2\rangle \langle q_2 1\rangle}+{\cal O}(\delta_2^0/\delta_1, \delta_1^0/\delta_2)~.
 \>
If we take the reverse consecutive limit, i.e. take leg $q_1$ soft and then leg $q_2$, the leading term in ${\tt CSL}(q_1^{+}, q_2^{+})$ is unchanged. However when the particles have different helicity, or when we extract sub-leading multi-soft terms, the order of limits will be important
and so we consider linear combinations by defining more general multi-soft limits
\<
\lim_{\alpha}\equiv(\alpha_{12\dots m} \lim_{\delta_m \to 0}\dots\lim_{\delta_2 \to 0}\lim_{\delta_1 \to 0} 
+\alpha_{21\dots m } \lim_{\delta_m \to 0}\dots \lim_{\delta_1 \to 0}\lim_{\delta_2 \to 0} +\dots)
\>
with the ellipses denoting all further permutations.
Thus we can define a multi-parameter family of 
 consecutive  soft limits
\<
{\tt \alpha CSL}_{n,1}( 1^{h_{q_1}},2^{h_{q_2}},\dots, m^{h_{q_m}}) A_n(1, \dots, n) =\lim_\alpha  {A}_{n+m}(\delta_1 q^{h_{q_1}}_1,\dots, \delta_m q^{h_{q_m}}_m, 1, \dots, n)~.
\>
 The effect of this general limit can be seen in the sub-leading term in the expansion of
 two soft, positive helicity gluons which a short calculation shows is given by
  \<
\left.  \lim_\alpha  {A}_{n+2}(\delta_1 q^{+}_1,  \delta_2 q^{+}_2,  \dots)\right|_{\rm sub-leading}&=&  \Big[
\frac{1}{\delta_1}\left(\frac{\alpha_{12} \vev{n q_2}\vev{q_1 1}+\alpha_{21}\vev{n1}\vev{q_1 q_2}}{\vev{q_1q_2}
  \vev{ q_1 1 }\vev{n q_1}\vev{q_2 1}}\right)[q_2 \partial_1]
\nn\\
  & &
  \kern-190pt 
+ \frac{\alpha_{12}}{\delta_1}\frac{[q_2 \partial_n]}{\vev{ n q_1 }\vev{q_1 q_2}}
  +\frac{1}{\delta_2}\left(\frac{\alpha_{12} \vev{n 1}\vev{q_1 q_2}+\alpha_{21}\vev{q_1 1}\vev{n q_2}}{\vev{q_1q_2}
  \vev{n q_2 }\vev{n q_1}\vev{q_2 1}}\right)[q_1 \partial_n]
  +\frac{\alpha_{21}}{\delta_2}\frac{[q_1 \partial_1]}{\vev{  q_2 1}\vev{q_1 q_2}} 
  \Big]A_n~.
  \>
 It is perhaps worthwhile to note that this expression is only valid for generic external momenta as we have neglected 
 holomorphic anomaly terms that can arise when external legs are collinear with soft legs, however it will turn out that they are not 
 necessary for our considerations. 
 
 For the case of mixed helicity the two orderings of limits already differ at leading order in the soft expansion and so we can 
 define a two parameter double-soft limit
 \<
 \label{eq:alpha_mix_hel}
  {\tt \alpha CSL}^{(0)}_{n,1}(q_1^{+}, q_2^{-})&=&\frac{1}{ \langle n q_1\rangle [q_2 1]}\left(\alpha_{21} \frac{ \langle n 1 \rangle}{[q_1q_2]}\frac{[q_1 1]}{\langle q_1 1\rangle}+\alpha_{12} \frac{[ n 1]}{\langle q_1q_2\rangle }\frac{\langle q_2\, n \rangle}{[
  q_2\, n]}\right)~, 
  \>
 where ${\tt \alpha CSL}^{(0)}$ has the overall $(\delta_{1}\delta_{2})^{-1}$ dependence stripped off. 
As described previously, we can  of course also consider the simultaneous double-soft limit
where both particles are taken soft together. 
The case of two gluons of the same helicity is identical to the consecutive limit but the case of 
one negative helicity and one positive is different. In four space-time dimensions the leading
order double-soft mixed helicity factor is given in the spinor helicity formalism by
\<
\label{DSLpm}
{\tt DSL}_{n,1}^{(0)}(q_1^{+},q_2^{-})&= & \frac{1}{\langle n |q_{12}|1]}\,
\bigg[ \frac{1}{2 p_{n}\cdot q_{12}}\, \frac{[n 1] \vev{n  q_2}^{3}}{\langle q_1q_2\rangle \langle n q_1\rangle } - \frac{1}{2 p_{1}\cdot q_{12}}\, 
\frac{\langle n   1\rangle [1q_1]^{3}}{[q_1q_2] [q_21]}
\bigg]\, , 
\>
where
\<
\qquad q_{12}:=q_{1}+q_{2} \, .
\>
This expression differs from the consecutive limit due to the sum of soft momenta
in the denominator but it is closest to the symmetric $\alpha_{12}=\alpha_{21}=\tfrac{1}{2}$ case.  
A similar ambiguity appears at sub-leading order when we consider the double-soft 
limit in the mixed helicity case. Explicit formulae can be found in \cite{Klose:2015xoa}
but in this case we will focus on the consecutive limits which can be calculated by 
repeated use of the single-soft limits \eqref{eq:single_soft}.

\section{Current algebra interpretation of Yang-Mills soft limits}

\begin{figure}
\begin{center}
\raisebox{0cm}{
\begin{minipage}{100pt}
\vspace{0cm}
\includegraphics[scale=0.4]{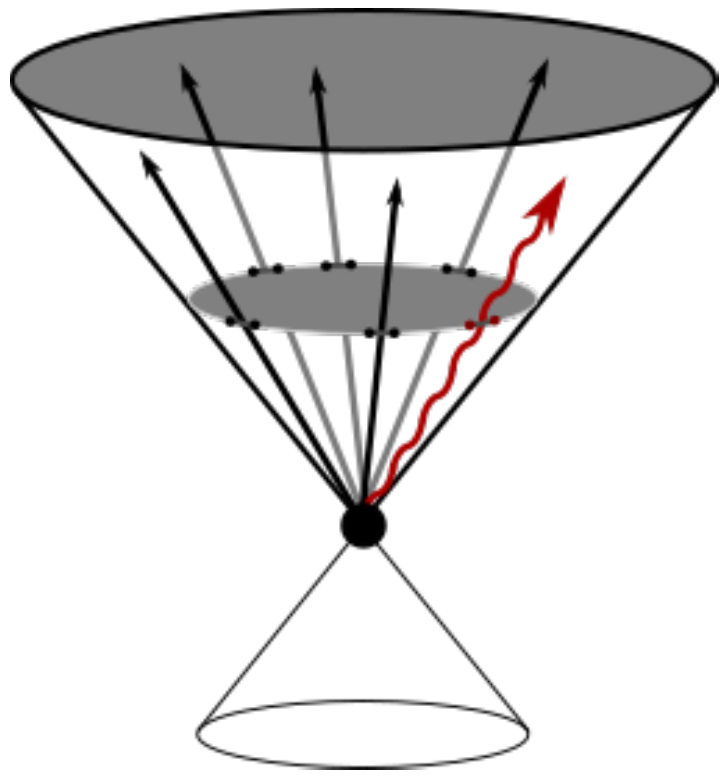}
\end{minipage} }
~$\longrightarrow$~~~
\raisebox{0cm}{
\begin{minipage}{120pt}
\vspace{0cm}
\includegraphics[scale=0.35]{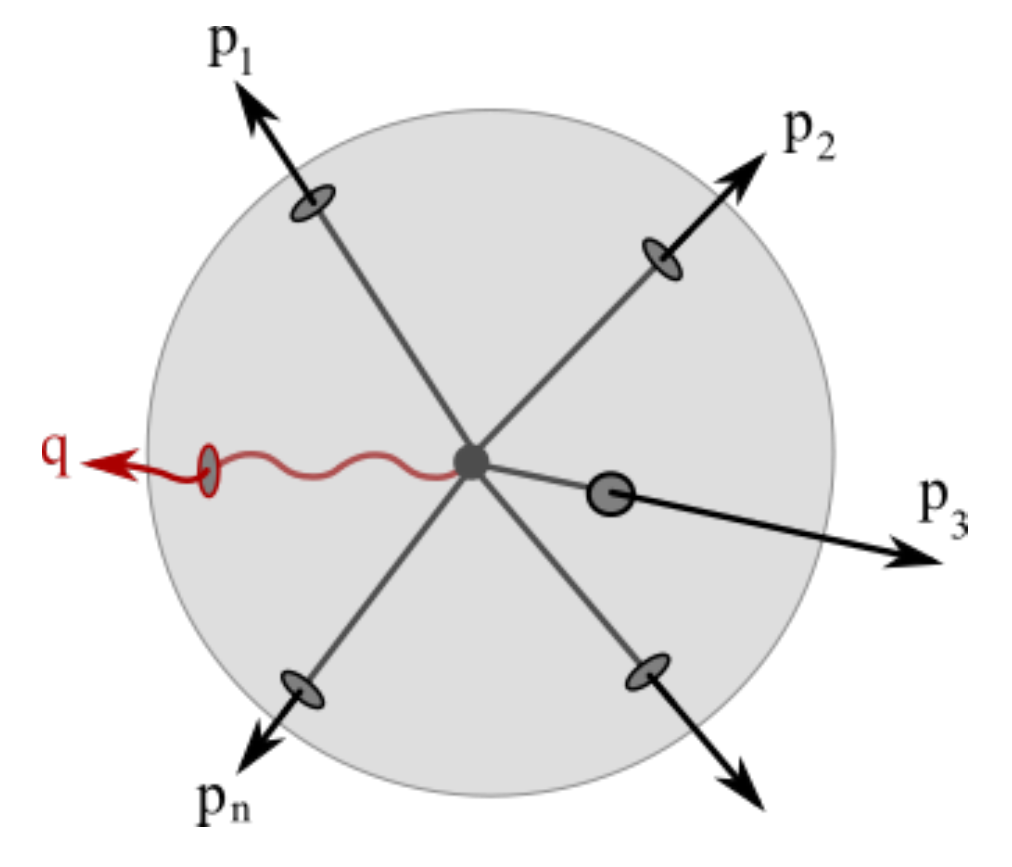}
\end{minipage} }
~$\longrightarrow$~~~
\raisebox{0cm}{\begin{minipage}{140pt}
\vspace{0cm}
\includegraphics[scale=0.35]{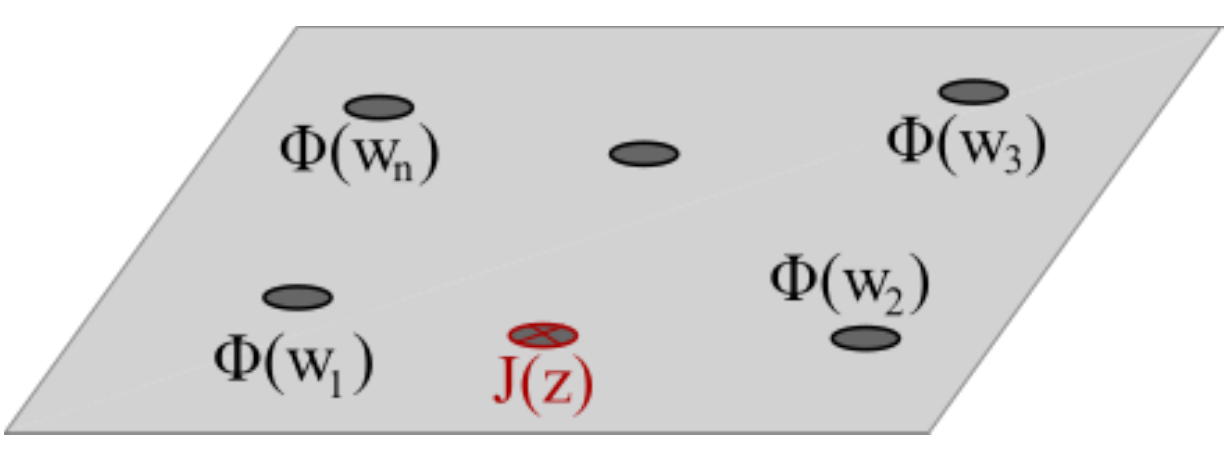}
\end{minipage}}
\caption{The gluons emerging from the amplitude travel on the light-cone and intersect with the sphere at null infinity, or any finite
time representation, at points which can be given complex coordinates $z$, for the soft gluon,  and $u_i$ for the hard gluons. The
soft gluon is then interpreted as the insertion of a current operator at the point $z$ while the $u_i$ denote the insertion
points of additional fields.}
\label{fig:sphere_gluon}
\end{center}
\end{figure}

Let us consider the single and double soft limits of Yang-Mills amplitudes but now instead of just 
focussing on colour-ordered partial amplitudes we examine the full amplitude
\<
{\cal A}_n(\{p_i, h_i, a_i\})=g_{\rm YM}^{n-2}\sum_{\sigma\in S_n/Z_n}A_n(\sigma_1, \dots, \sigma_n){\rm Tr}(T^{a_{\sigma_1}}\dots T^{a_{\sigma_n}})
\>
where the sum runs over all permutations, $S_n$, modulo those which are cyclic, $Z_n$, and $T^a$ are the generators of the colour
algebra which we will take to be $\alg{su}(N)$. We will drop the factors of $g_{\rm YM}$ but as we are only concerned 
with tree-level amplitudes they can be trivially restored. If we take the soft-limit of the $n+1$-particle amplitude we have
at leading order in the soft-expansion
\<
\left. \lim_{\delta \to 0}{\cal A}_{n+1}(\delta q_1,h_{q_1}, a; \{p_i, h_i, a_i\})\right|_{\tfrac{1}{\delta}}& =& 
\sum_{\sigma \in S_n} \frac{1}{\delta} S^{(0)}_{\sigma_n,\sigma_1} (q_1^{h_{q_1}}) 
A_n(\sigma_1, \dots,\sigma_n){\rm Tr}(T^{a} T^{a_{\sigma_1}}\dots T^{a_{\sigma_{n}}})~.\nn
\>
We can rewrite this soft limit in terms of two-dimensional position space variables by using the
parameterisation for the soft-gluon momentum
\<
\label{eq:coor_rep}
\lambda_{q_1}=\sqrt{\omega}(1, z)~, ~~~{\tilde \lambda}_{q_1}=\sqrt{\omega}(1, {\bar z})~,
\>
and for the hard momenta the parameterisation $\lambda_i=\sqrt{\nu_i}(1,u_i)$ and $\bar \lambda_i=\sqrt{\nu_i}(1,\bar u_i)$.
Essentially the complex $z$-variable describes the position of the intersection of the soft gluon's trajectory with the
sphere at asymptotic infinity 
and the $u_i$'s are the corresponding intersection points for the hard gluons, see 
\figref{fig:sphere_gluon}. We take all gluons to be outgoing and so this is in fact the anti-sky mapping. 
In these variables we have for the leading single soft factor  \eqref{eq:single_soft} , 
\<
(- \delta \omega )S^{(0)}_{i,j}(q_1^+)=\frac{1}{z-u_j}-\frac{1}{z-u_i}~,
\>
where we see that it has simple poles as the soft momenta approach the hard momentum in 
position space i.e. $z\rightarrow u_{i}$ for each $i$.
Now we define,
 \<
{\cal J}^a(z)\cdot {\cal A}_n(\{u_i, \bar u_i, \nu_i\})\equiv -\lim_{\delta \to 0} (\delta \omega){\cal A}_{n+1}(\delta p ,+, a; \{p_i, h_i, a_i\}),
\>
and so
\<
{\cal J}^a(z)\cdot {\cal A}_n(\{u_i, \bar u_i\})&=&
\sum_{\ell=1}^n
\Big[\frac{1}{z-u_{\ell}}\Big] \sum_{\tilde \sigma \in S_{n-1}}  A_n(\ell, \tilde \sigma_2, \dots, \tilde \sigma_n){\rm Tr}([T^a, T^{a_\ell}]T^{a_{\tilde \sigma_2}}\dots T^{a_{\tilde \sigma_n}})\nn\\
&=&\sum_{\ell=1}^n \frac{T_\ell^a}{z-u_\ell} {\cal A}_n(\{u_i, \bar u_i\})
\label{eq:singlesoftamplitude}
\>
where in the last line $T_\ell^a$ is understood to act (in the adjoint representation) only on the $\ell$-th particle colour factor in ${\cal A}_n$. This is the result of He \textit{et al} \cite{He:2015zea} 
who compared it to the OPE for a Ka{c}-Moody current $J^a$ and a primary field in a representation $R$
\<
J^a(z)\Phi_R^r(u)\sim \frac{(T^a_R)_s^r\Phi_R^s(u)}{z-u}
\>
and the consequent Ward Identity
\<
\vev{ J^a(z) \Phi_{R_1}(u_1,{\bar u}_1) \dots \Phi_{R_n}(u_n, {\bar u}_n)}=\sum^{n}_{\ell=1}\frac{T_{R_\ell}^a}{z-u_\ell}\vev{ \Phi_{R_1}(u_1,{\bar u}_1) \dots \Phi_{R_n}(u_n, {\bar u}_n)}~.
\label{eq:singlesoftCFT}
\>
Note that we use the notation $ \hJ^a $ in \eqref{eq:singlesoftamplitude} to denote the expression derived from a four-dimensional scattering amplitude perspective and the $ J^a $ in \eqref{eq:singlesoftCFT} to be the corresponding quantity from the two dimensional CFT perspective. It is the formal similarity of these two expressions that suggests the interpretation of the
soft-gluon limits as corresponding to the insertion of the current operator and it is interesting to ask to what extent this analogy can be extended. 

\subsection{An \texorpdfstring{$\alg{su}(N)$}{\textit{su(N)}} current algebra}
As a first step, it is natural to ask whether one can reproduce the current-current OPE
\<
J^a(z_1) J^b(z_2)\sim \frac{k \delta^{ab}}{(z_1-z_2)^2}+\frac{i f^{ab}{}_{c} J^c(z_2)}{z_1-z_2}~.
\>
At the level of correlation functions we would expect to find, 
\<
\label{eq:JJcorr}
\vev{J^a(z_1) J^b(z_2)\Phi(u_1) \dots } \sim \frac{k \delta^{ab}}{(z_1-z_2)^2}\vev{\Phi(u_1)\dots }+\frac{i f^{abc}}{z_1-z_2}\sum_{\ell}\frac{T^c_{\ell}}{z_2-u_\ell}\vev{\Phi(u_1)\dots }~.
\>
That the corresponding result can be found by analyzing the double-soft limits of amplitudes was shown in \cite{He:2015zea} and as it is
useful for our later results we rederive this fact in our notations.  
To this end we consider the double soft limit which we can write as 
\<
\label{eq:dbl_sft}
\lim_{\delta \to 0}{\cal A}_{n+2}(\delta_1 q_1,h_{q_1}, a_{q_1}; \delta_2 q_2,h_{q_2}, a_{q_2}; \{p_i, h_i, a_i\})&=&\nn\\
& &
\kern-240pt
  \sum_{\sigma \in S_n} \left\{ ({\tt CSL}_{\sigma_n,\sigma_1}( q_1^{h_{q_1}}, q_2^{h_{q_2}})+{\tt CSL}_{\sigma_n, \sigma_1}( q_2^{h_{q_2}}, q_1^{h_{q_1}}))
 {\rm Tr}(q_1 q_2 \sigma_1  \dots \sigma_{n})\right. \nn\\
& &
\kern-200pt +
 \sum_{i=2}^{n}S_{\sigma_n,\sigma_1}( q_1^{h_{q_1}})S_{\sigma_{i-1},\sigma_{i}}( q_2^{h_{q_2}})
   {\rm Tr}(q_1 \sigma_1 \dots \sigma_{i-1} q_2 \sigma_{i}  \dots \sigma_{n})\nn\\
& &  \left. \kern-150pt +
 i  {\tt CSL}_{\sigma_n,\sigma_1}( q_2^{h_{q_2}}, q_1^{h_{q_1}})f^{a_{q_2} a_{q_1}}{}_{c} ~{\rm Tr}(c \sigma_1  \dots \sigma_{n})\right\}
A_n(\sigma_1, \dots, \sigma_{n}) 
\>
in a slightly condensed notation with ${\rm Tr}(q_1q_2\sigma_1\dots\sigma_{n})={\rm Tr}(T^{a_{q_1}}T^{a_{q_2}}T^{a_{\sigma_1}}\dots T^{a_{\sigma_{n}}})$ etc. 
As we first want to find the OPE of two holomorphic 
currents we consider the leading order double soft limit of two positive helicity gluons.  
As the gluons have the same helicity the order
of limits is not relevant and so we simply take the
simultaneous limit with a single soft parameter $\delta_1=\delta_2=\delta$. 
We use the identity for consecutive soft-limits
\<
\label{eq:csl_pp_recomb}
{\tt CSL}^{(0)}_{\sigma_{n},\sigma_1}( q_1^{h_{q_1}},q_2^{h_{q_2}})+{\tt CSL}^{(0)}_{\sigma_{n},\sigma_1}( q_2^{h_{q_2}}, q_1^{h_{q_1}})= 
S^{(0)}_{\sigma_{n},\sigma_1}( q_1^{h_{q_1}})S^{(0)}_{\sigma_{n},\sigma_1}( q_2^{h_{q_2}})
\>
here with both helicities positive $h_{q_1}=h_{q_2}=+$, and so
we can combine the first two terms to find 
\<
\lim_{\delta \to 0}{\cal A}_{n+2}(\delta q_1,+, a_{q_1}; \delta q_2,+, a_{q_2}; \{p_i, h_i, a_i\})&=&\nn\\
& &\kern-140pt
  \sum_{\sigma \in S_n} \left\{ 
 \sum_{i=1}^{n}S^{(0)}_{\sigma_{n}, \sigma_1}(q_1^{+})S^{(0)}_{\sigma_{i-1}, \sigma_i}(q_2^{+})
   {\rm Tr}(q_1 \sigma_1 \dots \sigma_{i-1} q_2 \sigma_{i}  \dots \sigma_{n})
   \right. \nn\\
& &  \left. \kern-120pt 
+ i  {\tt CSL}^{(0)}_{\sigma_{n},\sigma_1}( q_2^{+}, q_1^{+})f^{a_{q_2} a_{q_1}}{}_{ c} ~{\rm Tr}(c \sigma_1  \dots \sigma_{n})\right\}
A_n(\sigma_1, \dots, \sigma_{n}) ~. 
\>
Re-writing the soft-factors in terms of polarisation vectors one recovers the result quoted in \cite{He:2015zea}. 

The singularities of these double soft factors in the collinear limit, as the soft momenta approach each other, can
be made clear by rewriting them in position space and examining the $z_1\rightarrow z_2$ limit. We can see
that there will be no singularities from the terms involving the products of single soft factors; these terms instead
give poles when the soft momenta 
approach the hard momenta. However the term involving the consecutive double soft limit does 
have such a singularity and gives rise to the non-trivial algebra between the holomorphic currents. More explicitly, defining 
\<
\label{eq:double_insert}
{\cal J}^a(z_1){\cal J}^b(z_2)\cdot {\cal A}_n(\{u_i, {\bar u}_i\})\equiv \lim_{\delta\to 0} (4~\delta^2~ \omega_{z_1}\omega_{z_2})
{\cal A}_{n+2}(\delta q_1, +, a; \delta q_2, +, b; \{p_i, h_i, a_i\})
\>
we find
\<
{\cal J}^a(z_1){\cal J}^b(z_2)\cdot {\cal A}_n(\{u_i, {\bar u}_i\})&\sim&\frac{1}{z_1-z_2}\sum_{\ell \neq m =1}^n\sum_{\sigma\in S_{n-2}} i f^{ba}{}_{c}\left(\frac{1}{z_2-u_\ell}-\frac{1}{z_2-u_m}\right)
\nn\\
& &\kern+150pt \times  {\rm Tr}(\ell c m \sigma_1\dots)A_n(\ell m \sigma_1\dots)\nn\\
&=&\frac{if^{ba}{}_{c}}{z_1-z_2}\sum_{\ell=1}^n\frac{1}{z_2-u_\ell}\sum_{\tilde \sigma \in S_{n-1}}{\rm Tr}([\ell, c]\tilde \sigma_1\dots)A_n(\ell \tilde\sigma_1\dots)\nn\\
&=&\frac{if^{ab}{}_{c}}{z_1-z_2}\sum_{\ell=1}^n\frac{T^c_\ell}{z_2-u_\ell}{\cal A}_n(\{u_i, {\bar u}_i\})~.
\>
Hence comparing with \eqref{eq:JJcorr} we see that the double soft-limit reproduces the structure of 
level-zero Kac-Moody or simply a current algebra as described in \cite{He:2015zea}.


\subsection{Sugawara construction for amplitudes}
If this is a correct interpretation, i.e. there is a level-zero Kac-Moody symmetry acting on amplitudes, we may try to construct
a holomorphic energy-momentum tensor via the Sugawara construction. Namely, we define
\<
T^S(z_1)=\gamma :J^aJ^a(z_1):~\equiv\gamma \lim_{z_2\to z_1}J^a(z_1)J^a(z_2)
\>
where $\gamma$ is a constant related to the dual Coxeter number, $h^{\vee}$, by $\gamma= \tfrac{1}{2 h^\vee}$, and the normal ordering is missing the usual 
singular term that occurs in the Kac-Moody case as the current algebra has no central extension. 
At the level of correlation functions this leads us to again consider double insertions of holomorphic currents and so, as the analogue for amplitudes is the double soft limit, we again consider \eqref{eq:double_insert} but now with contracted adjoint indices. That is we define 
\<
{\cal T}^S(z_1)=\tfrac{1}{2 h^\vee} \lim_{z_2\to z_1} {\cal J}^a{\cal J}^a(z_1)~.
\>
We could in principle work with a general Lie group but for simplicity we focus on the case of $\alg{su}(N)$ colour where we can contract
the indices $a$ and $b$ using  the identity
\<
(T^a)_i{}^j(T^a)_k{}^l=\delta^{l}_{i}\delta^j_k-\frac{1}{N}\delta^j_i\delta^l_k~.
\>
and in this case $h^\vee=N$.

To calculate the scattering amplitude analogue of the insertion of 
of an energy-momentum tensor into a correlation function we consider the singularities as
 either of the soft momenta approach one of the
hard momenta, say $u_m$, and then expand the $z_2$ dependence near $z_1$. 
Using the notation 
$z_{1m}=z_1-u_m$ for $m=1,\dots, n$ we write the result 
\<
\label{eq:T_insert}
{\cal T}^S(z_1)\cdot {\cal A}_n(\{u_i, {\bar u}_i\})&\equiv&
\frac{1}{2N}{\cal J}^a(z_1){\cal J}^a(z_1)\cdot {\cal A}_n(\{u_i, {\bar u}_i\})
\nn\\
&=&\sum_{m=1}^{n}\sum_{\sigma\in {S_{n-1}}}\Big\{\frac{1}{2N} \frac{1}{z_{1m}}\sum_{i=1}^{n-1} \frac{1}{z_{1\sigma_{i}}} {\rm Tr}([a,m]\sigma_1\dots[a, \sigma_i]\dots)\nn\\
& & \kern+40pt+\frac{1}{z^2_{1m}}{\rm Tr}(m\sigma_1\dots )
\Big\}A_n(m\sigma_1\dots)~.
\>
While this definition seems sensible we should ask whether it satisfies the basic properties of a
energy-momentum tensor for a conformal field theory. 

\subsection{Energy-momentum tensor/current OPE}
In two-dimensional field theory a defining characteristic
of $T^S(z_1)$ is its OPE with the holomorphic current operator
\<
T^S(z_1) J^a(z_3)\sim \frac{J^a(z_3)}{z_{13}^2}+\frac{\partial J^a(z_3)}{z_{13}}~.
\>
The analogous object for the amplitude is a specific case of the triple soft-limit
\<
\label{eq:trip_soft}
& &\lim_{\delta \to 0}{\cal A}_{n+3}(\delta q_1,h_{q_1}, a_{q_1}; \delta q_2,h_{q_2}, a_{q_2}; \delta q_3, h_{q_3}, a_{q_3} ; \{p_i, h_i, a_i\}) \\
& &= \sum_{\sigma \in S_n} \Big\{ 
 \sum_{s=1}^{n-1} \sum_{\substack{r=1 \\ r \neq s}}^{n}   S^{(0)}_{\sigma_{n},\sigma_1}(q_1^{h_{q_1}}) S^{(0)}_{\sigma_{s},\sigma_{s+1}}(q_2^{h_{q_2}})
S^{(0)}_{\sigma_{r},\sigma_{r+1}}(q_3^{h_{q_3}}){\rm Tr}(q_1 \dots \sigma_{s}q_2\dots\sigma_{r}q_3\dots)
\nn\\
& & +
\sum_{r=1}^{n-1}  \left(( 
{\tt CSL}^{(0)}_{\sigma_{n},\sigma_1}( q_1^{h_{q_1}}, q_2^{h_{q_2}}, )+
 {\tt CSL}^{(0)}_{\sigma_{n},\sigma_1}( q_2^{h_{q_2}}, q_1^{h_{q_1}})) S^{(0)}_{\sigma_r,\sigma_{r+1}}(q_3^{h_{q_3}})~{\rm Tr}(q_1q_2  \dots \sigma_{r}q_3\dots)\right. \nn\\
& &\left.+i 
{\tt CSL}^{(0)}_{\sigma_{n},\sigma_1}( q_2^{h_{q_2}}, q_1^{h_{q_1}})
S^{(0)}_{\sigma_r,\sigma_{r+1}}(q_3^{h_{q_3}})f^{a_{q_2}a_{q_1}}{}_{c}
{\rm Tr}(c\dots \sigma_r q_3\dots)+{\rm (q_2\leftrightarrow q_3)}+{\rm (q_1\leftrightarrow q_3)} \right)\nn\\
& & +\left(
{\tt CSL}^{(0)}_{\sigma_{n},\sigma_1}( 1^{h_1}, 2^{h_2},3^{h_3})
  {\rm Tr}(1 2 3\sigma_1  \dots \sigma_{n})+{\rm perm( 1,2,3)}\right)
 \Big\}
A_n(\sigma_1, \dots, \sigma_{n})\nn ~. 
\>
We take all gluons to have positive helicity, such that the triple and double soft limits
are the same regardless of whether we take the consecutive or simultaneous soft limit
and so, as in the double-soft case, we set all soft parameters equal. 
The product of three holomorphic currents is then defined to be 
\<
\label{eq:triple_insert}
{\cal J}^a(z_1){\cal J}^b(z_2){\cal J}^c(z_3) \cdot {\cal A}_n(\{u_i, {\bar u}_i\})& &\equiv  \\
& &\kern-150pt- \lim_{\delta\to 0} (\delta^2~ \omega_{z_1}\omega_{z_2}\omega_{z_3})
{\cal A}_{n+3}(\delta q_1, +, a, \delta q_2, +, b, \delta q_3, +, c; \{p_i, h_i, a_i\})~.\nn
\>
In the standard two-dimensional field theory calculation, to compute the OPE of the energy-momentum tensor with the current operator we collect 
the terms which are singular as $z_3$ approaches either $z_1$ or $z_2$. In the soft-limit such terms correspond
to the soft momentum $q_3$ becoming collinear with either $q_1$ or $q_2$. In particular the
contributions from the triple soft terms split into two groups, those where the soft particles $q_1$ and $q_2$ are adjacent which come 
with a colour factor $(N-\tfrac{1}{N}){\rm Tr}(c\sigma_1 \dots \sigma_n)$ while those where $q_1$ and $q_2$ are split by particle 
$q_3$ have $(-\tfrac{1}{N}){\rm Tr}(c\sigma_1 \dots \sigma_n)$. Combining these terms we find that the $(-\tfrac{1}{N})$ factors
cancel. There are further contributions from terms involving double soft limits times single soft factors which have colour 
structures $f^{c a}{}_{d}{\rm Tr}(a \dots \sigma_r d\dots) $ however these terms cancel between themselves. This makes use 
of the identity \eqref{eq:csl_pp_recomb} as well as the fact that
\<
{\rm Tr}(T^{a} T^{a_{\sigma_{1}}}  \dots T^{a_{\sigma_r}}[T^{c},T^a]\dots)+
{\rm Tr}([T^{c},T^a] T^{a_{\sigma_{1}}}  \dots T^{a_{\sigma_r}}T^{a} \dots)=0~.
\>
Following the conformal 
field theory calculation, we then expand this in powers of $z_{21}=z_2-z_1$ and keep the leading term (in principle there could be a more 
singular term however this term is absent which corresponds to the level of the current algebra being zero ). We finally have 
\<
{\cal T}^S(z_1){\cal J}^c(z_3)
\cdot {\cal A}_n(\{u_i, {\bar u}_i\})&\sim&
\frac{1}{z_{31}^2}\sum_{\sigma\in S_{n}} \left(\frac{1}{z_1-u_{\sigma_1}}-\frac{1}{z_1-u_{\sigma_n}}\right)
  {\rm Tr}(c  \sigma_1\dots\sigma_n)A_n( \sigma_1\dots\sigma_n)\nn\\
&=&\frac{1}{z_{31}^2}\sum_{\ell=1}^n\frac{1}{z_1-u_\ell}\sum_{\tilde \sigma \in S_{n-1}}{\rm Tr}([c, \ell]\tilde \sigma_1\dots)A_n(\ell \tilde\sigma_1\dots)\nn\\
&=&\frac{1}{z^2_{31}}{\cal J}^c(z_1)
\cdot {\cal A}_n(\{u_i, {\bar u}_i\})
\>
which after expanding the $z_1$ dependence on the right hand side is the required result. 


\subsection{Energy-momentum OPE}
\label{sec:TTOPE}
%
%
Using the OPE of holomorphic currents it can be shown that 
the Sugawara energy-momentum tensor for a current algebra ($k=0$)
satisfies the OPE
\<
\label{eq:TT_OPE}
T(z_1)T(z_3)\sim \frac{2T(z_3)}{(z_1-z_3)^2}+\frac{\partial T(z_3)}{(z_1-z_3)}
\>
and so defines a conformal field theory with vanishing central charge. 
Correspondingly, inserted into a correlator of primary fields $\Phi(u_i)$, we have that
\<
\langle T(z_1)T(z_3)\Phi(u_1)\dots \Phi(u_n)\rangle &=&
\Big[\frac{2}{(z_1-z_3)^2}+\frac{\partial_{z_3}}{(z_1-z_3)}\\
& &
\kern-45pt +
\sum_{i=1}^n\Big\{\frac{\Delta}{(z_1-u_i)^2}+\frac{\partial_{u_i}}{(z_1-u_i)}\Big\}\Big]\times
\langle T(z_3)\Phi(u_1)\dots \Phi(u_n)\rangle~. \nn
\>
In terms of the soft-limits of amplitudes this Ward identity should be reproduced by the collinear part of the quadruple soft limit for positive helicity gluons.
 Once again as we are 
only considering positive helicity gluons there is no ambiguity in the order of limits. In the two-dimensional field theory
construction one starts with
pairs of currents, $J^{a_1}(z_1)J^{a_2}(z_2)$ and $J^{a_3}(z_3)J^{a_4}(z_4)$, and 
to calculate the OPE one
first extracts the terms which become singular as either $z_1$ or $z_2$ approaches either
$z_3$ or $z_4$ and then takes the limits $z_2\to z_1$ and $z_4 \to z_3$. For the amplitude one 
similarly considers those terms which are singular as either $q_1$ or $q_2$ become collinear with 
$q_3$ or $q_4$. 
The singular terms get contributions from quadruple soft terms, where all soft particles are adjacent, 
from triple-soft terms and from the ``double'' double-soft terms. 
Carefully combining all these terms we find, after taking the $z_2 \to z_1$ limit, that the singular terms in 
 the quadruple soft limit are
\<
\label{eq:TT_soft}
{\cal T}^S(z_1){\cal T}^S(z_3)\cdot {\cal A}&\equiv &\lim_{\delta \to 0}~\frac{(\prod_{i=1}^4  \omega_{z_i})}{(2N)^2}  {\cal A}_{n+4}(\delta q_1,+, a; \delta q_1,+, a; \delta q_3, +, b ;\delta q_3, +, b; \{p_i, h_i, a_i\})\nn \\
& &\kern-110pt \sim 
\frac{2}{z_{13}^2 (2N)} 
\sum_{\sigma \in S_n}
\sum_{i=1}^n 
\left(\frac{u_{\sigma_n}-u_{\sigma_1}}{(z_{3}-u_{\sigma_n})(z_3-u_{\sigma_1})}\right)
\left(\frac{u_{\sigma_{i-1}}-u_{\sigma_i}}{(z_{1}-u_{\sigma_{i-1}})(z_1-u_{\sigma_i})}\right)
{\rm Tr}(a\sigma_1\dots a \sigma_i
\dots)A_n~.\nn
\>
This can be compared with the definition of a single insertion 
of the energy-momentum tensor in \eqref{eq:T_insert} and, after further expanding
the $z_1$ dependence near $z_3$,  we
see that we indeed reproduce the OPE in \eqref{eq:TT_OPE} as expected. 

While we restrict our considerations to gluon amplitudes it is interesting to note
that the sub-leading soft theorem for a graviton has been identified 
with the Ward identity for a two-dimensional energy-momentum tensor \cite{Kapec:2016jld}
which suggests interpreting the collinear limit of double-soft gluons as a graviton. This is reminiscent of recent work \cite{Stieberger:2014cea, Stieberger:2015qja, Stieberger:2015kia,Nandan:2016ohb} showing that amplitudes describing the interactions
of gravitons with gluons can be written  as linear combinations of amplitudes in
which the graviton is replaced by a pair of collinear gluons.

\subsection{Comments on a Knizhnik-Zamolodchikov equation}
Given the above construction for the energy-momentum tensor it is interesting to ask if we can derive conformal Ward 
identities by considering correlation functions with insertions of the energy-momentum tensor.  However, \textit{a priori},
even if there is a sensible conformal field theory interpretation of the  asymptotic states the current algebra may 
only form part of it, which is to say the full energy-momentum tensor $T(z)$ would be given by 
\<
T(z)=T'(z)+T^S(z)~.
\>
If we consider the individual terms in the mode expansion
\<
({\hat L}_{-n}\Phi)(u)&=&\oint \frac{dz}{2\pi i} \frac{1}{(z-u)^{n-1}} T(z) \Phi(u)~,  \nn\\
({\hat J}^a_{-n}\Phi)(u)&=&\oint \frac{dz}{2\pi i} \frac{1}{(z-u)^{n}} J^a(z) \Phi(u)
\>
we have that 
\<
{\hat L}_{-m}={\hat L}'_{-m}+\gamma \sum_{l\leq -1} {\hat J}^a_{l}{\hat J}^a_{m-l}+\gamma \sum_{l> -1} {\hat J}^a_{m-l}{\hat J}^a_{-l}~.
\>
In conformal field theory we can insert this inside a correlation functions of primary fields and in particular if we consider the case where 
$T'(z)$ is absent we can derive the Knizhnik-Zamolodchikov equation. 
\<
\vev{\Phi(u_1)\dots ({\hat L}_{-1}\Phi)(u_m)\dots \Phi(u_n)}&=&\nn\\
& &\kern-40pt \vev{\Phi(u_1)\dots(\gamma(\sum_{l\leq -1} {\hat J}^a_{l}{\hat J}^a_{-l-1}+\sum_{l> -1} {\hat J}^a_{-l-1}{\hat J}^a_{-l})\Phi)(u_m)\dots
\Phi(u_n)}.\nn
\>
Turning now to the scattering amplitudes, we consider the individual terms in the expression for an insertion of the energy-momentum tensor following from 
the double-soft limit, \eqref{eq:T_insert}. We have at leading order
\<
\oint_{{\cal C}_{u_m}} \frac{dz_1}{2\pi i} ~z_{1m} {\cal T}^S(z_1)\cdot {\cal A}_n(\{u_i, {\bar u}_i\})&=& {\cal A}_n(\{u_i, {\bar u}_i\})
\>
where ${\cal C}_{u_m}$ is a contour surrounding the point $u_m$. This has the expected form for an insertion of the energy-momentum tensor into 
a correlation function of primary fields with weight one. More non-trivially the sub-leading term is given by
\<
\oint_{{\cal C}_{u_m}} \frac{dz_1}{2\pi i}~ {\cal T}^S(z_1)\cdot {\cal A}_n(\{u_i, {\bar u}_i\})&= &\frac{1}{N}\sum_{\substack{\ell=1\\ \ell\neq m}}^{n}\frac{T_m^a\otimes T^a_{\ell}}{u_m-u_\ell}{\cal A}_n(\{u_i, {\bar u}_i\})~.
\>
The usual conformal Ward identity would then imply 
\<
\label{eq:KZ_diff}
\partial_{u_m}{\cal A}_n(\{u_i, {\bar u}_i\})={\cal L}'_{-1}\cdot {\cal A}_n(\{u_i, {\bar u}_i\})+\frac{1}{N}\sum_{ \substack{\ell=1 \\ \ell \neq m} }^{n}\frac{T_m^a\otimes T^a_{\ell}}{u_m-u_\ell}{\cal A}_n(\{u_i, {\bar u}_i\})
\>
where ${\cal L}'_{-1}$ is the residue of $T'(z_1)$ at $z_1=u_m$. It is tempting to ask what would happen if the ${\cal L}'_{-1}$ term would be absent, that is, 
if the current algebra described the complete CFT.  As an example one can consider the MHV amplitudes. It is useful to define a
rescaled quantity $G_n$ as follows
\<
G_n(\{u_i, {\bar u}_i, h_i\})=\prod_{\ell=1}^n \nu_\ell^{h_\ell} {\cal A}_n(\{u_i, {\bar u}_i\})~.
\>
In the simplest three-particle case 
\<
G_3^{MHV}(1^-,2^-,3^+)= \frac{ (u_1-u_2)^3}{(u_2-u_3)(u_3-u_1)} ({\rm Tr}(T^{a_1}T^{a_2}T^{a_3})-{\rm Tr}(T^{a_1}T^{a_3}T^{a_2}))
\>
and it is immediately obvious that \eqref{eq:KZ_diff} with ${\cal L}'_{-1}$ absent is satisfied for $u_3$ but not $u_1$ or $u_2$. 
In fact it is easy to check explicitly that the KZ-equation with ${\cal L}'_{-1}$  absent continues to hold for the positive helicity gluons  
 in higher point MHV amplitudes (currently done with Mathematica to seven-points).  It is slightly non-trivial to check that the double trace terms, 
 which arise when $m$ and $\ell$ in \eqref{eq:KZ_diff} are not adjacent in a specific colour-ordered term, vanish. This requires that the colour ordered amplitudes
 satisfy a number of  identities. For example,  for the $m=3$ KZ equation one such relation at four points is the vanishing of the ${\rm Tr}(T^{a_1}T^{a_4}){\rm Tr}(T^{a_2}T^{a_3})$
 double-trace terms, which requires that 
  \<
 & &\frac{A(1,2,3,4)}{u_3-u_1}+ \frac{A(1,4,3,2)}{u_3-u_1}- \frac{A(1,2,3,4)}{u_3-u_2} - \frac{A(1,3,2,4)}{u_3-u_2}
 \nn\\
& &- \frac{A(1,4,2,3)}{u_3-u_2}- \frac{A(1,4,3,2)}{u_3-u_2}+ \frac{A(1,3,2,4)}{u_3-u_4}+ \frac{A(1,4,2,3)}{u_3-u_4}=0~.
\label{amplituderelation4point}
\>
More generally for the $m$-th KZ-equation the vanishing of the 
\<
& &{\rm Tr}(T^{a_m} T^{a_{\sigma_1} } \dots T^{a_{\sigma_{\ell-1} } }){\rm Tr}(T^{a_{\sigma_{\ell} }}\dots T^{a_{\sigma_{n-1}} })
\>
term requires 
\<
& &\kern-60pt \sum_{{\rm cyc} \{\sigma_{\ell}, \dots, \sigma_{n-1}\}}\left( \Big[\frac{1}{u_{m}-u_{\sigma_\ell}}-\frac{1}{u_{m}-u_{\sigma_{\ell-1}}}\Big]A_n(m\sigma_1\dots \sigma_\ell \dots \sigma_{n-1})
\right. \nn\\
& &\kern+60pt \left. -\Big[\frac{1}{u_{m}-u_{\sigma_1}}-\frac{1}{u_{m}-u_{\sigma_{n-1}}}\Big]A_n(m, \sigma_{\ell}\dots \sigma_{n-1}\sigma_{1}\dots \sigma_{\ell-1})\right)=0
\>
where $\rm{cyc} \{\sigma_{\ell}, \dots, \sigma_{n-1}\}$ denotes the sum over all cyclic permutations. 
We checked all such identities hold up to seven points for MHV amplitudes. However, starting with the six-point NMHV amplitude it no longer
 appears that the KZ equations holds even for the positive helicity gluons. This suggests, unsurprisingly,  that 
 if there is a CFT interpretation there is additional structure and as there is further universal 
 behaviour in the soft-limits we attempt to also interpret these in the context of a two-dimensional description.

Let us round off this section with an interesting connection of the above amplitude relations with the so called Bern-Carrasco-Johansson (BCJ)\cite{Bern:2008qj} relations resulting from the color-kinematic duality of gauge theory amplitudes. We can re-write the previous example of the four point amplitude relation \eqref{amplituderelation4point} in a different way using Kleiss-Kuiff and the photon decoupling relations, which gives $ A(1,2,3,4)=A(1,4,3,2) $ and $ A(1,3,2,4)=A(1,4,2,3) $. Using these relations we can rewrite the left-hand side of \eqref{amplituderelation4point} as,
\beqa
A(1,2,3,4)(\frac{1}{u_{31}}-\frac{1}{u_{32}})+A(1,4,2,3)(\frac{1}{u_{34}}-\frac{1}{u_{32}})=0\nn\\
A(1,2,3,4)\frac{u_{12}}{u_{31}}+A(1,4,2,3)\frac{u_{42}}{u_{34}}=0.
\label{amplituderelation4point-02}
\eeqa
Now using the representations of the spinor helicity variables $ \langle ij\rangle =\sqrt {\nu_i\nu_j}u_{ij}$ and $ [ij] =\sqrt {\nu_i\nu_j}\bar{u}_{ij}$ and four-point momentum conservation we get,
\beq
\frac{u_{12}u_{34}}{u_{31}u_{24}}=\frac{\langle 12\rangle \langle 34\rangle[12]}{\langle 24\rangle \langle 31\rangle [12]}=\frac{s_{12}}{s_{24}},
\eeq
where $ s_{ij}=(p_i+p_j)^2 $ are the Mandelstam invariants. Hence, using these change of variables \eqref{amplituderelation4point-02} and resultantly \eqref{amplituderelation4point} becomes,
\beq
A(1,2,3,4)s_{12}=s_{13}A(1,4,2,3),
\label{4pointBCJ}
\eeq
which is remarkably the $ 4-$point BCJ relation. Thus the MHV amplitude relation derived from demanding the vanishing of the double-trace terms in our KZ equations leads to the BCJ relation at least for $ 4$ points. This connection is harder to see for higher number of points and we will report on this in a future publication.

\section{Sub-leading currents}
\label{sec:sub}

So far we have only been considering the leading soft terms, however it is known that the sub-leading soft 
terms for Yang-Mills amplitudes are also universal. This behaviour, 
the gluon version of the Low theorem \cite{Low:1954kd, GellMann:1954kc, Low:1958sn, Burnett:1967km,  Gross:1968in, Jackiw:1968zza}
described in \secref{sec:ss}, is understood to be valid only 
at tree-level and for generic configurations of the remaining external hard legs \cite{Larkoski:2014bxa}. We will only consider
such configurations and so neglect boundary terms in the kinematic space where 
additional particles become collinear. 
At loop level such corrections could no longer be avoided and so it would be interesting to carefully understand
these terms, and more generally to make contact with the SCET \cite{Bauer:2000ew, Bauer:2000yr, Bauer:2001ct, Bauer:2001yt} description used in \cite{Larkoski:2014bxa}. 
 We thus repeat the above analysis at the sub-leading order
by defining the sub-leading currents, ${\cal J}^a_{\rm sub}$, as
\<
{\cal J}_{\rm sub}^a(z;\omega_z)\cdot {\cal A}_n(\{u_i, \bar u_i, \nu_i\})\equiv 
-\lim_{\delta \to 0}  (1+\delta \p_\delta)\omega_z {\cal A}_{n+1}(\delta q ,+, a; \{p_i, h_i, a_i\})
\>
where the $(1+\delta \p_\delta)$ factor picks out the sub-leading soft term. We have that 
\<
{\cal J}_{\rm sub}^a(z;\omega_z)\cdot {\cal A}_n(\{u_i, \bar u_i,\nu_i\})
&=&\sum_{\ell=1}^n \frac{\big[ \omega_z \partial_{\nu_\ell} + \tfrac{\omega_z}{\nu_\ell}(\bar{ z}-\bar{u}_\ell)\bar \partial_\ell\big]T_\ell^a}{z-u_\ell} {\cal A}_n(\{u_i, \bar u_i, \nu_i\})
\>
where $\bar \partial_\ell=\tfrac{ \partial}{\partial \bar z_\ell}$ and $\partial_{\omega_\ell}=\tfrac{ \partial}{\partial \omega_\ell}$. If we again attempt to interpret this in terms of 
the OPE of a current with a primary field we must consider the primary fields as depending on an auxiliary
variable $\nu$ and we have 
\<
J_{\rm sub}^a(z;\omega)\Phi_R^r(w; \nu)\sim \frac{ (T^a_R)^r_s  \big[\omega \partial_\nu+ \tfrac{\omega }{\nu }(\bar{ z}-\bar{u})\bar \partial\big] \Phi_R^s(w;\nu)}{z-u}~.
\>

The appearance of an auxilliary parameter and derivatives in the representation 
 is (very vaguely) reminiscent of  the Cartan element of the non-compact affine $\widehat{\alg{sl}_2}$ current algebra in the principle continuous 
series representations, $\Phi^{j,\ell}(w)$, for which we introduce the complex parameter $x$, and the OPE is given by 
\<
J^3(z) \Phi^{j,\ell}(w;x)& \sim & \frac{t^3 \Phi^{j,\ell}(w;x)}{z-u}+\frac{k\ell}{2}\Phi^{j,\ell}(w;x)
\>
where $t^3=x\tfrac{\partial}{\partial x}$. 
In the case at hand we see that, in addition to having a one-parameter family of such currents, there is no extension as $k=0$. 
 Adamo and Casali \cite{Adamo:2015fwa} have, in their world-sheet
theory with null infinity as its target space, given the definition of a charge with reproduces the sub-leading soft factor. It would be interesting to understand if there is a relation between the currents introduced here and their charges which have the interpretation as rotating the space of null generators. It is also 
known that these sub-leading soft terms in four-dimensions
are intimately connected to the conformal invariance of the theory \cite{Larkoski:2014bxa}.

\subsection{Sub-leading algebra}
\label{sec:subalg}

To continue the two-dimensional interpretation of the collinear soft limits to sub-leading order and given these new currents we analyse the algebra that they form with the leading currents and amongst themselves.  

\paragraph{$J_{\rm sub} J$ :}
To find the OPE of the Kac-Moody current $J^a(z)$ with $J_{\rm sub}^a(z)$
we must consider part of the double soft limit at sub-leading order. The order of limits in this case is relevant and
a specific prescription must be given with the results dependent on the prescription. We  start with the consecutive limit 
and take leg $q_2$ soft before leg $q_1$:
\<
\lim_{\delta_1 \to 0}\lim_{\delta_2 \to 0} (1+\delta_1\partial_{\delta_1}) \delta_2 
{\cal A}_{n+2}(\delta_1 q_1,h_{q_1}, a_{q_1}; \delta_2 q_2,h_{q_2}, a_{q_2}; \{p_i, h_i, a_i\})&=&\nn\\
& &
\kern-340pt
  \sum_{\sigma \in S_n} \left\{\Big[
  S^{(0)}_{q_1,\sigma_1}( q_2^{h_{q_2}})S^{(1)}_{\sigma_{n},\sigma_1}(q_1^{h_{q_1}})
  + S^{(0)}_{\sigma_{n},q_1}(q_2^{h_{q_2}})S^{(1)}_{\sigma_{n},\sigma_1}(q_1^{h_{q_1}})\Big]
 {\rm Tr}(q_1 q_2 \sigma_1  \dots \sigma_{n})\right. \nn\\
& &
\kern-300pt +
 \sum_{i=2}^{n}S^{(0)}_{\sigma_{i-1},\sigma_i}( q_2^{h_{q_2}}) S^{(1)}_{\sigma_{n},\sigma_1}( q_1^{h_{q_1}})
   {\rm Tr}(q_1 \sigma_1 \dots \sigma_{i-1} q_2 \sigma_{i}  \dots \sigma_{n})\nn\\
& &  \left. \kern-300pt +
 i  S^{(0)}_{\sigma_{n},q_1}(q_2^{h_{q_2}} )S^{(1)}_{\sigma_{n},\sigma_1}(q_1^{h_{q_1}}) f^{a_{q_2} a_{q_1}}{}_{ c} ~{\rm Tr}(c \sigma_1  \dots \sigma_{n})\right\}
A_n(\sigma_1, \dots, \sigma_{n}) ~.
\>
An important feature of this formula is that, as previously, the singularities as $q_1$ and $q_2$ become collinear 
only occur in the third set of terms on the right-hand side as those that appear in the first set of terms cancel amongst themselves
\<
\lim_{\delta_1 \to 0}\lim_{\delta_2 \to 0} (1+\delta_1\partial_{\delta_1}) \delta_2  {\cal A}_{n+2}(\delta_1 q_1,h_{q_1}, a_{q_1}; \delta_2 q_2,h_{q_2}, a_{q_2}; \{p_i, h_i, a_i\})&=&\nn\\
& &
\kern-340pt
-  \sum_{\sigma \in S_n} \left\{\Big[ \frac{\vev{\sigma_{n} \sigma_1}}{\vev{\sigma_{n} q_2}\vev{q_2 \sigma_1}} 
  \left(\frac{[q_1 \pb_{\sigma_1}]}{\vev{q_1\sigma_1}}+\frac{[q_1 \pb_{\sigma_{n}}]}{\vev{\sigma_{n} q_1}} \right)\Big]
 {\rm Tr}(q_1 q_2 \sigma_1  \dots \sigma_{n})\right. \\
& &
\kern-300pt +\dots \nn\\
& &  \left. \kern-300pt +
 i  \frac{\vev{\sigma_{n} q_1 }}{\vev{\sigma_{n} q_2}\vev{q_2 q_1}} 
  \left(\frac{[q_1 \pb_{\sigma_1}]}{\vev{q_1\sigma_1}}+\frac{[q_1 \pb_{\sigma_{n}}]}{\vev{\sigma_{n} q_1}} \right)
   f^{a_{q_2} a_{q_1}}{}_{ c} ~{\rm Tr}(c \sigma_1  \dots \sigma_{n})\right\}
A_n(\sigma_1, \dots, \sigma_{n}) ~.\nn
\>
Using the coordinate representation \eqref{eq:coor_rep} we can find the most singular part as $z_1\to z_2$
\<
\label{eq:sing_sub_1}
\lim_{\delta_1 \to 0}\lim_{\delta_2 \to 0} (1+\delta_1\partial_{\delta_1}) \delta_2 \omega_{z_1} \omega_{z_2} {\cal A}_{n+2}&\sim &
\frac{if^{a_{q_1}a_{q_2}}{}_{c}}{z_{12}} \sum_{\ell=1}^{n}
 \Big[
 \frac{  T^c_\ell \omega_{z_1} (\p_{ \nu_{\ell } }+\tfrac{\bar z_2-\bar u_\ell}{\nu_\ell }\bar \p_{\ell})}{z_2-u_{\ell }}
 \Big]
{\cal A}_n~.
\>
If we had instead taken $\delta_2$ to zero
after $\delta_1$ we would find 
\<
\lim_{\delta_2 \to 0}\lim_{\delta_1 \to 0} (1+\delta_1\partial_{\delta_1}) \delta_2 
{\cal A}_{n+2}(\delta_1 q_1,h_{q_1}, a_{q_1}; \delta_2 q_2,h_{q_2}, a_{q_2}; \{p_i, h_i, a_i\})&=&\nn\\
& &
\kern-340pt
  \sum_{\sigma \in S_n} \left\{\Big[
S^{(1)}_{\sigma_{n},q_2}(q_1^{h_{q_1}})  S^{(0)}_{\sigma_{n},\sigma_1}(q_2^{h_{q_2}})+ S^{(1)}_{q_2,\sigma_1}(q_1^{h_{q_1}} )S^{(0)}_{\sigma_{n},\sigma_1}(q_2^{h_{q_2}})\Big]
 {\rm Tr}(q_1 q_2 \sigma_1  \dots \sigma_{n})\right. \nn\\
& &  \left. \kern-300pt +\dots +
 i  S^{(1)}_{q_2,\sigma_1}(q_1^{h_{q_1}} )S^{(0)}_{\sigma_{n},\sigma_1}(q_2^{h_{q_2}}) f^{a_{q_2} a_{q_1}}{}_{ c} ~{\rm Tr}(c \sigma_1  \dots \sigma_{n})\right\}
A_n(\sigma_1, \dots, \sigma_{n}) ~.
\>
Again one can show that in the collinear limit the potentially singular terms only come from the last line above, 
however in this case there are in fact no non-vanishing singular terms in the collinear limit.
Such terms potentially could have arisen from the holomorphic anomaly
\<
\frac{\p}{\p\bar \lambda^{\dot a}}\frac{1}{\vev{\lambda\mu}}=\pi \epsilon_{\dot a\dot b}\bar \mu^{\dot b}\delta^{(2)} (\vev{\lambda\mu})
\>
which gives 
\<
\lim_{\delta_2 \to 0}\lim_{\delta_1 \to 0} (1+\delta_1\partial_{\delta_1}) \delta_2 {\cal A}_{n+2}&=&\nn\\
& &\kern-150pt
 - \sum_{\sigma \in S_n} \left\{\Big[\dots +i\pi f^{a_{q_2}a_{q_1}}{}_{c}\frac{\vev{\sigma_{n}\sigma_1}}{\vev{q_2 q_1}}
  \left(-\frac{[q_1\sigma_{n}]}{\vev{q_2 \sigma_1}}\delta^{(2)}(\vev{\sigma_{n} q_2})\right. \right.\nn\\
  & &\kern-50pt \left. \left.+\frac{[q_1\sigma_{1}]}{\vev{q_2 \sigma_{n}}}\delta^{(2)}(\vev{ q_2\sigma_1})
  \right){\rm Tr}(c \sigma_1  \dots \sigma_{n})\right\}
A_n(\sigma_1, \dots, \sigma_{n})~.
\>
However these terms are also of order one when $q_1$ becomes collinear with $q_2$. 
Thus taking a combination with parameters $\alpha_{12}$ and $\alpha_{21}$ 
only the $\alpha_{21}$ parameter contributes and from the definition 
\<
\label{eq:subalg_mixed_def}
{\cal J}_{\rm sub}^a(z_1){\cal J}^b(z_2)\cdot {\cal A}_n(\{u_i, {\bar u}_i\})& &\equiv  \\
& &\kern-100pt \lim_{\alpha} (1+\delta_1\partial_{\delta_1}) \delta_2~ \omega_{z_1} \omega_{z_2}
{\cal A}_{n+2}(\delta_1 q_1, +, a; \delta_2 q_2, +, b; \{p_i, h_i, a_i\})~\nn
\>
we find
\<
\label{eq:subalg_mixed}
{\cal J}_{\rm sub}^a(z_1){\cal J}^b(z_2)\cdot {\cal A}_n(\{u_i, {\bar u}_i,\omega_i\})& =&
\frac{i\alpha_{21} f^{ab}{}_{c}}{z_{12}} \sum_{\ell=3}^{n+2}
 \Big[
 \frac{ T^c_\ell \omega_{z_1} ( \p_{ \nu_{\ell } }+\tfrac{\bar z_2-\bar u_\ell}{\nu_\ell}\bar\p_\ell)  }{z_2-u_{\ell }}
 \Big]
{\cal A}_n~\nn
\>
or in the notation of the OPE, after a slight rearrangement involving exchanging the soft legs 
and renaming the parameter $\jjs$, 
\<
{J}^a(z_1) {J}_{\rm sub}^b(z_2;\omega_{z_2})\sim \frac{i \jjs  f^{ab}{}_{c} J_{\rm sub}^c(z_2;\omega_{z_2})}{z_{12}}~.
\>
Note that if we choose $\jjs$ to vanish then the OPE between the
leading soft current and the sub-leading current vanishes and these sectors decouple as the 
parameter $\alpha_{12}$ doesn't appear. For brevity we will make the choice to set 
$\jjs=1$ and $\jsj=0$, however it can always be reintroduced. 
Using this OPE,  one can compute the OPE of the Sugawara energy-momentum tensor with
${J}_{\rm sub}^a(z;\omega_{z})$ 
\<
T^S(z_1) {J}_{\rm sub}^a(z_2;\omega_{z_2})
\sim \frac{{J}_{\rm sub}^a(z_2;\omega_{z_2})}{z_{12}^2}
+ \frac{\partial {J}_{\rm sub}^a(z_2;\omega_{z_2})}{z_{12}}~,
\>
where we have not included the composite term $f^{a d c} :\!J^a J^d_{\rm sub}\!:(z_2,\omega_{z_2})$ which appears at $1/z_{12}$. 
This is what one would expect for a field of weight one. 

\paragraph{$J_{\rm sub} J_{\rm sub}$ :} To find the OPE of the sub-leading currents with themselves we must calculate the sub-leading behaviour of each soft particle in the limit of two soft gluons
\<
\lim_{\alpha} (1+\delta_1\partial_{\delta_1}) (1+\delta_2\partial_{\delta_2}) 
{\cal A}_{n+2}(\delta_1 q_1, +, a; \delta_2 q_2, +, b; \{p_i, h_i, a_i\})~\nn\\
& &\kern-200pt=
\sum_{\sigma \in S_n} \left\{\Big[
 \dots +
 i  \frac{f^{bac}}{\vev{q_2 q_1}}\left(
 \frac{ \alpha_{21}[q_2\pb_{\sigma_{n} }]}{\vev{\sigma_{n} q_1} }+\frac{\alpha_{21}[q_2 \pb_{\sigma_{1}}]}{\vev{q_1\sigma_1}}\right.\right.
 \nn\\
 & & \left. \left.\kern-200pt +
 \frac{\alpha_{12} [q_1\pb_{\sigma_{n} }]}{\vev{\sigma_{n} q_2} }+\frac{\alpha_{12}[q_1 \pb_{\sigma_{1}}]}{\vev{q_2\sigma_1}}
 \right)  ~{\rm Tr}(c \sigma_3  \dots \sigma_{n+2})\right\}
A_n(\sigma_1, \dots, \sigma_{n}) ~\nn
\>
where we have included contributions from both orderings and from which we have 
\<
{\cal J}_{\rm sub}^a(z_1;\omega_{z_1}){\cal J}_{\rm sub} ^b(z_2;\omega_{z_2})\cdot {\cal A}_n(\{u_i, {\bar u}_i\})
&=&\nn\\
&  &\kern-60pt
\frac{if^{ab}{}_{c}}{z_{12}} \sum_{\ell=1}^{n}
 \Big[
 \frac{ T^c_\ell (\alpha_{21} \omega_{z_1} +\alpha_{12}  \omega_{z_2})(\p_{ \nu_{\ell } }+\tfrac{(\bar z_2-\bar u_\ell)}{\nu_\ell})\bar\p_\ell  }{z_2-u_{\ell }}
 \Big]
{\cal A}_n~\nn
\>
or in the notation of the OPE and writing the parameters as $\jsjsab$ and $\jsjsba$ we have 
\<
{J}_{\rm sub}^a(z_1;\omega_{z_1}){J}_{\rm sub}^b(z_2;\omega_{z_2})\sim \frac{i f^{ab}{}_{c} J_{\rm sub}^c(z_2;\jsjsba  \omega_{z_1}+\jsjsab \omega_{z_2})}{z_{12}}~.
\>
Here we see an example where there doesn't appear to be a natural choice for the ordering of the soft limits and so we simply keep both and parameterise the ambiguity 
by $\jsjsab$ and $\jsjsba$.

Given our previous considerations and the form of the OPE, one might attempt to repeat the Sugawara construction for these sub-leading currents. 
While we don't analyse the general case, at least in the symmetric
case where $\jsjsab=\jsjsba=1$ using the OPE it is straightforward to show that one can define an operator for each value of $\omega_{z_1}$
\<
{T}^S_{\rm sub}(z_1, \omega_{z_1})=\frac{1}{2N}{J}_{\rm sub}^a(z_1; \omega_{z_1}){J}_{\rm sub}^a(z_1;-\omega_{z_1})
\>
which acts like a  sub-leading energy-momentum tensor in that it satisfies
\<
{T}^S_{\rm sub}(z_1,\omega_{z_1}) {J}_{\rm sub}^a(z_2;\omega_{z_2})\sim \frac{{J}_{\rm sub}^a(z_2;\omega_{z_2})}{z_{12}^2}
+ \frac{\partial {J}_{\rm sub}^a(z_2;\omega_{z_2})}{z_{12}}~.
\>


\section{Anti-holomorphic currents}

We can of course repeat the previous calculations for the negative helicity gluon 
and find the anti-holomorphic currents. We define 
 \<
\bar {\cal J}^a(z)\cdot {\cal A}_n(\{u_i, \bar u_i, \nu_i\})\equiv \lim_{\delta \to 0} (\delta \omega){\cal A}_{n+1}(\delta p ,-, a; \{p_i, h_i, a_i\}),
\>
so that the current still acts with the adjoint action but because of the missing minus sign it acts from the right rather than the left or, alternatively,
with the complex conjugate generator
\<
\bar {\cal J}^a(z)\cdot {\cal A}_n(\{u_i, \bar u_i\})
&=&\sum_{\ell=1}^n \frac{\bar T_\ell^a}{\bar z-\bar u_\ell} {\cal A}_n(\{u_i, \bar u_i\})~.
\>

\subsection{OPE for anti-holomorphic currents}
Slightly more non-trivially we can consider
the mixed helicity double soft limit and attempt to reproduce the action of the holomorphic 
currents on the anti-holomorphic discussed in \cite{He:2015zea}. We can again start from \eqref{eq:dbl_sft}
however now $h_{q_1}=+$ while $h_{q_2}=-$. In this case there is an ambiguity in 
the double soft limit which we parameterise in case of  consecutive limits by
using a specific case of the general multi-parameter limit
\<
\lim_{\alpha}\equiv(\alpha_{12} \lim_{\delta_2 \to 0}\lim_{\delta_1 \to 0} 
+\alpha_{21} \lim_{\delta_1 \to 0}\lim_{\delta_2 \to 0})~.
\>
The corresponding two-parameter family of 
 consecutive double-soft factors, ${\tt \alpha CSL}^{(0)}_{\sigma_{n},\sigma_1}$,
still satisfies the identity analogous to \eqref{eq:csl_pp_recomb}
\<
\label{eq:scsl_pm_recomb}
{\tt \alpha CSL}^{(0)}_{\sigma_{n},\sigma_1}( 1^{+},2^-)+{\tt \alpha CSL}^{(0)}_{\sigma_{n},\sigma_1}( 2^{-}, 1^+)= 
S^{(0)}_{\sigma_{n},\sigma_3}( 1^+) S^{(0)}_{\sigma_{n},\sigma_3}( 2^-)~.
\>
We thus find that the terms singular in $z_{12}$ arise from the ${\tt \alpha CSL}^{(0)}_{\sigma_{n},\sigma_1}( 2^{-},1^+)$ 
term. More explicitly, if we start from
\<
\label{eq:ahol_mixed}
{\cal J}^a(z_1)\bar{{\cal J}}^b(z_2)\cdot {\cal A}_n(\{u_i, {\bar u}_i\})\equiv  -
\lim_\alpha \delta_1 \delta_2~ \omega_{z_1} \omega_{z_2}
{\cal A}_{n+2}(\delta_1 q_1, +, a; \delta_2 q_2, -, b; \{p_i, h_i, a_i\})~\nn
\>
then we find that 
\<
{\cal J}^a(z_1)\bar{{\cal J}}^b(z_2)\cdot {\cal A}_n(\{u_i, {\bar u}_i\})& &\kern-15pt=i f^{ab}{}_{c}
\sum_{\ell=1}^{n}
 \Big[\dots +
\alpha_{12}   \frac{ \bar T^c_\ell   }{z_{12}(\bar z_2-\bar u_{\ell })}
- \alpha_{21}  \frac{  T^c_\ell   }{\bar z_{12}( z_2- u_{\ell })}
\nn
\\& & \kern+20pt + \alpha_{21} \frac{z_{12}}{\bar z_{12}} \frac{T^c_\ell}{(z_2-z_\ell)^2}
 +\dots\Big]
{\cal A}_n~,\nn
\>
where we have included the sub-leading term that has a non-trivial phase as $z_1$ encircles $z_2$. 
This expression can be interpreted as a non-trivial OPE between the holomorphic and anti-holomorphic 
currents of the form 
\<
\label{eqn:JJb_OPE}
{J}^a(z_1)\bar{{J}}^b(z_2)\sim if^{ab}{}_{c}\Big[\jjb \frac{\bar{J}^c(z_2)}{z_{12}}-\jbj \frac{{J}^c(z_2)}{\bar {z}_{1 2}}
-\jbj  \frac{z_{12}}{\bar {z}_{12}}\p J^c(z_2)\Big] ~, 
\>
where the OPE parameters are related to the order of the soft limits by $\jjb=\alpha_{12}$ and $\jbj=\alpha_{21}$.
This structure for the OPE between the holomorphic and anti-holomorphic currents appears similar 
to that found in the CFTs describing supergroup coset models considered in \cite{Ashok:2009xx}. If we require this OPE to be consistent 
with the complex conjugation $(J^a(z_1))^\ast=\bar J^a(z_1)$ we find the constraint $\jjb^\ast=\jbj$
and $\jbj^\ast=\jjb$. For the soft-limit which naturally implies real parameters this requires $d_1=d_2$
which is to say the symmetric choice of parameters. 

It is interesting to compare this with what one finds in the simultaneous double-soft
by computing
\<
-\lim_{\delta\to 0}\delta^2~ \omega_{z_1} \omega_{z_2}
{\cal A}_{n+2}(\delta q_1, +, a, \delta q_2, -, b; \{p_i, h_i, a_i\})~,\nn
\>
and making use of the results of \cite{Klose:2015xoa, Volovich:2015yoa}.
Focusing on the singular terms this gives 
\<
\frac{i f^{ab}{}_{c}}{(\omega_{z_1}+\omega_{z_2})^2}
\sum_{\ell=1}^{n}
 \Big[\dots 
 + \frac{ \bar T^c_\ell   \omega_{z_2}^2 }{z_{12}(\bar z_2-\bar u_{\ell })}
-   \frac{  T^c_\ell  \omega_{z_1}^2 }{\bar z_{12}( z_2- u_{\ell })}
 +\dots\Big]
{\cal A}_n~.\nn
\>
With the previous interpretation of the leading order soft limits in terms of currents $J^a$ and $\bar J^a$
this expression doesn't appear to make sense due to the appearance of the soft-particle energies 
$\omega_{z_i}$ on the right-hand side. One could of course attempt to introduce 
a family of currents parametrized by $\omega$ already at leading order;
alternatively we can additionally demand that $\omega_{z_1}=\omega_{z_2}$ in which case we
reproduce the consecutive answer with a symmetric choice for the parameters $\jjb$ and $\jbj$. 

\paragraph{$ \bar J J_{\rm sub}$ :}
To complete the algebra we must compute the OPE of the anti-holomorphic current with the sub-leading current.
The calculations are essentially the same as those above and we again start from
\<
\label{eq:sub_ahol_mixed}
\bar{{\cal J}}^a(z_1) {\cal J}_{\rm sub}^b(z_2;\omega_{z_2})\cdot {\cal A}_n(\{u_i, {\bar u}_i\})\equiv  
\nn\\
& & \kern-60pt 
-\lim_\alpha \delta_1(1+\delta_2 \partial_{\delta_2})~ \omega_{z_1} \omega_{z_2}
{\cal A}_{n+2}(\delta_1 q_1, -, a, \delta_2 q_2, +, b; \{p_i, h_i, a_i\})~\nn
\>
which implies 
\<
\label{eq:amp_JbJsub}
\bar{{\cal J}}^a(z_1) {\cal J}_{\rm sub}^b(z_2;\omega_{z_2})\cdot {\cal A}_n(\{u_i, {\bar u}_i\})& &\kern-15pt=i f^{ab}{}_{c}
\sum_{\ell=1}^{n}
 \Big[\dots -
2 \frac{\alpha_{21}}{z_{12} }   \frac{  \tfrac{ \omega_{z_2}}{\omega_{z_1}} \bar T^c_\ell  }{(\bar z_2-\bar u_{\ell })}
+
3 \alpha_{21}\frac{\bar z_{12} }{z_{12} }   \frac{  \tfrac{ \omega_{z_2}}{\omega_{z_1}} \bar T^c_\ell  }{ (\bar z_2-\bar u_{\ell })^2}
\nn\\
& &\kern+80pt
- \frac{\alpha_{12} }{\bar z_{12}} \frac{ T^c_\ell \omega_{z_2} \big[ \partial_{\omega_\ell}+\tfrac{\bar z_2-\bar u_\ell }{\nu_\ell}\bar \p_\ell \big] }{( z_2- u_{\ell })}
 +\dots\Big]
{\cal A}_n~.\nn
\>
As it stands this doesn't appear to be interpretable as an OPE between 
an anti-holomorphic current and a sub-leading current depending on the parameter
$\omega_{z_2}$ due to the explicit appearance of $\omega_{z_1}$ on the right-hand
side. Thus we are lead to imposing a particular ordering
for the soft limits where we take the particle corresponding to the sub-leading current
to be soft after the leading order current, that is $\alpha_{21}=0$ and $\alpha_{12}=1$.
This, taking the sub-leading limit after the leading limit,  is the same as in the holomorphic
sector and as there seems a reasonable choice. 
This corresponds to an OPE between the non-holomorphic 
currents and the sub-leading holomorphic current 
\<
\bar{{J}}^a(z_1) {J}_{\rm sub}^b(z_2;\omega_{z_2})  \sim - if^{ab}{}_{c}
 \frac{  {J}_{\rm sub}^c(z_2;\omega_{z_2})}{\bar z_{12} }~.
\>

In some aspects this lack of choice is unappealing and it would be interesting to understand it better.
As a small step in this direction one can again consider the simultaneous
double soft limit at sub-leading order the expressions for which can be found in \cite{Klose:2015xoa}. In this case the simultaneous limit is not the same as the 
consecutive limit and for the case of mixed helicity is not given by
products of single soft limits. The simultaneous limit mixes the terms where
particle $q_1$ and $q_2$ are sub-leading but by focussing on the terms with anti-holomorphic derivatives one can identify those terms corresponding to the
sub-leading terms in the soft-expansion for the positive helicity gluon. 
Denoting the simultaneous double-soft factor ${\tt DSL}_{n,1}(q_1^{h_1},q_2^{h_2})$ 
the relevant, singular, sub-leading terms are
\<
\left. {\tt DSL}_{n,1}(q_1^-,q_2^+)\right|_{\rm singular}
=\frac{[nq_2]^2 [q_2 \tilde \partial_1]}{[n q_1][q_1 q_2] \langle 1|q_{12}|n]}
-\frac{[nq_2]^2 [q_2 \tilde \partial_n]}{[n q_1][q_1 q_2] (2 p_n\cdot q_{12})}~.
\>
This prescription does not include the contact terms, which involve no derivatives, 
\<
{\tt DSL}^{(1)}_{n,1}(q_1^{-},q_2^{+},)\vert_{\rm contact}&=& \frac{[n\,q_2]^2 \vev{q_1\, n}}{[n\, q_1]}\frac{1}{(2p_{n}\cdot q_{12})^2}+\frac{\vev{1q_1}^2 [q_21]}{\vev{1q_2}}\frac{1}{(2p_1\cdot q_{12})^2}.
\>
however as they are not singular in the collinear limit this is no loss. 
Expanding the collinear limit, gives the analogous result to 
\eqref{eq:amp_JbJsub} 
\<
i f^{ab}_c \sum_{\ell=1}^{n}
 \Big[\dots 
- \frac{1 }{\bar z_{12}} \frac{ T^c_\ell \tfrac{\omega^2_{z_2}}{\omega_{z_1}+\omega_{z_2}} \big[ \partial_{\omega_\ell}+\tfrac{\bar z_2-\bar u_\ell }{\nu_\ell}\bar \p_\ell \big] }{( z_2- u_{\ell })}
 +\dots\Big]
{\cal A}_n~.\nn
\>
Here we see that the troublesome non-derivative terms in \eqref{eq:amp_JbJsub} don't appear. 
 Of course, as before, due to the non-local nature of the simultaneous limit
we find the soft-particle energies entering as $\tfrac{\omega^2_{z_2}}{\omega_{z_1}+\omega_{z_2}}$ and so the simultaneous limit, by itself, does
not give a good prescription. As this limit involves the sub-leading soft-terms this behaviour may be related to 
failure of the Low theorem in general. Understanding these terms better will be essential if there is to be any 
progress at loop-level where despite the breakdown of the Low theorem there is some evidence of 
universal behaviour \cite{Brandhuber:2015vhm} of a restricted type and in particular the SCET framework is useful in trying to better understand these terms
\cite{Larkoski:2014bxa}. 

\paragraph{$J_{\rm sub} \bar J_{\rm sub}$ :}
Finally we consider the OPE of a sub-leading current with its anti-holomorphic 
analogue. This is related to the sub-sub-leading double-soft limit. While such sub-sub-leading behaviour has not been studied in the 
simultaneous double-soft limit it
is straightforward to define using consecutive single-soft limits though naturally 
this again leads to the introduction of parameters. Starting from 
\<
\label{eq:subsub_ahol_mixed}
{{\cal J}}_{\rm sub}^a(z_1;\omega_{z_1}) \bar {\cal J}_{\rm sub}^b(z_2;\omega_{z_2})\cdot {\cal A}_n(\{u_i, {\bar u}_i\})& \equiv  &\nn\\
& &\kern-160pt-
\lim_\alpha (1+\delta_2 \partial_{\delta_2})(1+\delta_2 \partial_{\delta_2})~ \omega_{z_1} \omega_{z_2}
{\cal A}_{n+2}(\delta_1 q_1, +, a; \delta_2 q_2, -, b; \{p_i, h_i, a_i\})~\nn
\>
and so 
\<
{{\cal J}}_{\rm sub}^a(z_1;\omega_{z_1}) \bar {\cal J}_{\rm sub}^b(z_2;\omega_{z_2})
\cdot {\cal A}_n(\{u_i, {\bar u}_i\})& &\kern-15pt=i f^{ab}{}_{c}
\sum_{\ell=1}^{n}
 \Big[\dots 
- \frac{\alpha_{12} }{z_{12}} \frac{ \bar T^c_\ell \omega_{z_1} \big[ \partial_{\omega_\ell}+\tfrac{z_2-u_\ell}{\nu_\ell} \p_\ell \big] }{( \bar z_2- \bar u_{\ell })}\nn\\
& &\kern-80pt
- \frac{\alpha_{12}\bar z_{12} }{z_{12}} \frac{ \bar T^c_\ell \omega_{z_1} \big[ \partial_{\omega_\ell}+\tfrac{z_2-u_\ell}{\nu_\ell} \p_\ell \big] }{( \bar z_2- \bar u_{\ell })^2}
+ \frac{\alpha_{21} }{\bar z_{12}} \frac{ T^c_\ell \omega_{z_2} \big[ \partial_{\omega_\ell}+\tfrac{\bar z_2-\bar u_\ell}{\nu_\ell} \pb_\ell \big] }{( z_2-  u_{\ell })}
\nn\\
& &
-2 \frac{\alpha_{21} z_{12} }{\bar z_{12}} \frac{ T^c_\ell \omega_{z_2} \big[ \partial_{\omega_\ell}+\tfrac{\bar z_2-\bar u_\ell}{\nu_\ell} \pb_\ell \big] }{(  z_2-  u_{\ell })^2}
 +\dots\Big]
{\cal A}_n~\nn
\>
which corresponds to the OPE 
\<
{{J}}_{\rm sub}^a(z_1;\omega_{z_1}) \bar{J}_{\rm sub}^b(z_2;\omega_{z_2}) \sim if^{ab}{}_{c}\Big[
\frac{ {J}_{\rm sub}^c(z_2; \jsbjs  \omega_{z_2})}{\bar z_{12} }
-\frac{ \bar{J}_{\rm sub}^c(z_2; \jsjsb  \omega_{z_1})}{z_{12} }\nn\\
+\frac{  \bar z_{12}}{z_{12} }\pb \bar{J}_{\rm sub}^c(z_2; \jsjsb \omega_{z_1})
+2 \frac{  z_{12}}{\bar z_{12} } \p {J}_{\rm sub}^c(z_2; \jsbjs  \omega_{z_2}) \Big]~.
\>

\subsection{CFT interpretation of anti-holomorphic currents}
\label{sec:JbCFT}
It is interesting to  compute the OPE of the anti-holomorphic current with the holomorphic energy-momentum
tensor found via the Sugawara construction. This can be done directly by using the formulae above to compute 
the OPE of two holomorphic currents with an anti-holomorphic current and then extracting the singular terms
as the two holomorphic currents approach each other. 
At leading order in $z_{13}$ we find
\<
J^a(z_1) J^a(z_2) \bar J^c(z_3)\sim f^{ca}{}_d f^{ad}{}_{e}\Big[ c_1 \frac{\bar J^e(z_3)}{z_{13}^2}+\Big[c'_2  \frac{z_{21}}{\bar z_{21}}\frac{1}{z_{13}^2}+
c'_3\frac{\bar  z_{21}}{ z_{21}}\frac{1}{\bar z_{13}^2}\Big]J^e(z_3)\Big]~,
\>
where the constants $c_1$ and $c'_2, c'_3$ are related to the parameters in the OPEs of $J^a$ with $J^a$ \eqref{eq:JJ} and $J^a$ with $\bar J^a$ \eqref{eq:JJbar}. 
The terms less singular in $1/z_{13}$ can be computed and are more complicated but one can already see at this leading order
 the unwanted appearance of $J^a(z_3)$ in the OPE; however all such terms, while non-vanishing as $z_2\to z_1$,
 can be identified by their phase. If we define the energy-momentum tensor by the contour integral 
 \<
 T(z_1)=\frac{1}{4 \pi i h^\vee} \oint \frac{dz_2}{z_{21}} J^a(z_1) J^a(z_2)
 \>
 where $h^\vee$ is again the dual Coxeter number then we can drop all unwanted terms 
 depending on $z_{21}$ and $\bar z_{21}$ so that we have the result
 \<
  T(z_1) \bar J^c(z_3)\sim  c_1 \frac{\bar J^c(z_3)}{z_{13}^2}+c_1\frac{\bar z_{13}}{z_{13}^2}\pb \bar J^c(z_3)+\dots
  \>
 where we have now included some of the sub-leading terms and the coefficient $c_1$ is given by $c_1=d_1^2$. 
 The remaining sub-leading terms are either composite
 operators of the form $f^{ca}{}_{d} :J^a\bar J^d:(z_3)$ or of the form $\pb J^c$, $\p \bar J^c$. These later terms
 would be related to composite terms of the former type if we imposed the Maurer-Cartan 
 \<
 \pb J^c-\p \bar J^c-i f^{c}{}_{ad} J^a\bar J^d =0
 \>
 and current conservation 
 \<
  \pb J^c+\p \bar J^c =0
 \>
 equations. However we restrict our attention to the leading singularity terms for the present. 
 
One can alternatively start from the triple soft-limit of 
\eqref{eq:trip_soft} but where we now take the third helicity to be negative, $h_{q_3}=-$. 
We are interested in the case where we trace over the colour indices $a_{q_1}$ and $a_{q_2}$, 
extracting the terms that are singular as $z_3$ approaches $z_1$ or $z_2$ and 
then taking the limit $z_2\to z_1$.  For example, and again taking the gauge group to be $\alg{su}(N)$, 
the terms that potentially contribute are 
\<
\label{eq:trip_soft_ahol}
& & \lim_{\delta_1 \to 0}\lim_{\delta_2 \to 0}\lim_{\delta_3 \to 0}\sum_a {\cal A}_{n+3}(\delta_1 q_1,+, a; \delta_2 q_2,+, a; \delta_3 q_3, -, b ; \{p_i, h_i, a_i\})\nn\\
& &= \sum_{\sigma \in S_n} \Big\{\dots+i \sum_{r=1}^{n-1}
   \left(
 {\tt CSL}^{(0)}_{\sigma_{n},\sigma_1}( q_3^{-}, q_1^{+})S^{(0)}_{\sigma_r,\sigma_{r+1}}(q_2^{+})\right.\nn
 \\
& &~~~~~~~~~~\left.- {\tt CSL}^{(0)}_{\sigma_{n},\sigma_1}( q_3^{-}, q_2^{+})S^{(0)}_{\sigma_r,\sigma_{r+1}}(q_1^{+})\right)
f^{b a}{}_{ c}{\rm Tr}(a \dots \sigma_r c \dots)
\nn\\
& & ~~~~~~~~~~+(N-\tfrac{1}{N})
\left(
{\tt CSL}^{(0)}_{\sigma_{n},\sigma_1}( q_1^{+}, q_2^{+}, q_3^{-})+{\tt CSL}^{(0)}_{\sigma_{n},\sigma_1}( q_2^{+}, q_1^{+}, q_3^{-})
\right. \nn\\
& &~~~~~~~~~~
\left. +{\tt CSL}^{(0)}_{\sigma_{n}, \sigma_1}(q_3^{-}, q_1^{+}, q_2^{+})+{\tt CSL}^{(0)}_{\sigma_{n},\sigma_1}( q_3^{-}, q_2^{+}, q_1^{+}) \right) {\rm Tr}(b \sigma_1  \dots)
\nn\\
& &~~~~~~~~~~
-\tfrac{1}{N}  \left(
{\tt CSL}^{(0)}_{\sigma_{n}, \sigma_1}( q_1^{+}, q_3^{-}, q_2^{+})+{\tt CSL}^{(0)}_{\sigma_{n},\sigma_4}( q_2^{+}, q_3^{-}, q_1^{+})\right)
 {\rm Tr}(b \sigma_1  \dots) \Big\}\nn\\
 & &~~~~~~~~~~\times
A_n(\sigma_1, \dots, \sigma_{n}) ~. 
\label{eq:triplesoft-masterformula}
\>
However it remains to specify exactly how to take the multi-soft limit and the result depends heavily on the
prescription. We could use the general multi-parameter soft limit
\<
\lim_\alpha =\alpha_{123} \lim_{\delta_3\to 0} \lim_{\delta_2\to 0} \lim_{\delta_1\to 0}+
\alpha_{321} \lim_{\delta_1\to 0} \lim_{\delta_2\to 0} \lim_{\delta_3\to 0}\dots
\>
however it is simplified by symmetrizing the order in which the positive helicity gluons are taken
soft i.e. we take $\alpha_{123}=\alpha_{213}$, $\alpha_{132}=\alpha_{231}$
and $\alpha_{312}=\alpha_{321}$. In the above expression one can see two different colour structures:
those with Tr$(b \sigma_1\dots)$ and those with $f^{b a}{}_{c}{\rm Tr}(a\dots \sigma_r c\dots)$. Focusing on
the former we can show that the terms with coefficient $\tfrac{1}{N}$ cancel while the remaining 
terms give, here only including terms of order $\tfrac{1}{z_{13}^2}$ and $\tfrac{1}{\bar z_{13}^2}$, 
\<
N \sum_{\sigma \in S_n} \Big\{&\dots& -\Big[
4\alpha_{123}\frac{1}{z_{13}^2}\frac{1}{\bar z_3-\bar u_{\sigma_1}} 
+2\alpha_{132}
\frac{z_{21}}{z_{13}^2\bar z_{21}}\frac{1}{z_3 -  u_{\sigma_1}}
\\
& &
+2(2 \alpha_{312}+\alpha_{132})
\frac{\bar z_{21}}{\bar z_{13}^2  z_{21}}\frac{1}{z_3 -  u_{\sigma_1}}\Big]{\rm Tr}([b,\sigma_1] \sigma_2\dots)+ \dots\Big\}A_{n}(\sigma_1\dots)~.\nn
\>
Using the prescription to drop terms with non-trivial monodromy as $z_2$ circles $z_1$ this multi-soft limit can be written as
\<
{\mathcal T}(z_1){\ahJ}^a(z_3)\sim 2\alpha_{123} \frac{ \ahJ^a(z_3) }{z_{13}^2 }
+ 2\alpha_{123} \frac{\bar z_{13}  }{z_{13}^2 }\pb \ahJ^a(z_3)+\dots
\label{eq:TJbarCSL}
\>
where we have here included the local term at the sub-leading $\tfrac{\bar z_{13}}{z^2_{13}}$ order. We can see that we reproduce at
$\tfrac{1}{z^2_{13}}$-order the structure following from the OPE calculation if we make the appropriate choice for the ordering
of the consecutive soft-limits.
There are additionally terms at order
$\tfrac{1}{z_{13}}$ which have a bi-local structure and in this case the terms with colour structure
$f^{ba}{}_{c}{\rm Tr}(a\dots \sigma_r c\dots)$ in \eqref{eq:trip_soft_ahol} do not vanish but instead also have
a bi-local form. We have not carefully matched these terms with those appearing in 
the OPE.

To compare this result to the one obtained from the simultaneous triple soft limit we use the  above formula \eqref{eq:triplesoft-masterformula} and specifically focus on the third and fourth lines since those are the only singular contributions in the collinear limit. 
In \cite{Volovich:2015yoa} the simultaneous triple-soft formulae with three adjacent soft gluons of mixed helicities were derived 
for the following cases 
\beqa
A_{n+3}(\delta_1 q_1^+,  \delta_2 q_2^-, \delta_3 q_3^-, 1, \ldots,n)\big{|}_{\delta_1 \sim \delta_2 \sim \delta_3 \rightarrow 0} \rightarrow \mathcal{S}^{+--}_{n,1}A_{n}  \,\nn\\
A_{n+3}(\delta_1 q_1^+, \delta_2 q_2^-, \delta_3 q_3^+, 1, \ldots,n)\big{|}_{\delta_1 \sim \delta_2 \sim \delta_3 \rightarrow 0} \rightarrow  \mathcal{S}^{+-+}_{n,1}A_{n}~.
\label{tsl0pmm}
\eeqa
We can obtain the other possible triple-soft terms needed for the different permutations in \eqref{eq:triplesoft-masterformula} from \eqref{tsl0pmm} by conjugation, i.e. flipping bra to ket and vice-versa., hence
\beq
\mathcal{S}^{-+-}_{n,1}=\mathcal{S}^{+-+}_{n,1}\vert^* \quad\text{and}\quad \mathcal{S}^{-++}_{n,1}=\mathcal{S}^{+--}_{n,1}\vert^* ,
\eeq
and,  to get the remaining configurations, by exchanging the neighbouring labels $ n $ and $ 1 $, i.e.
\beq
\mathcal{S}^{++-}_{n,1}=\mathcal{S}^{-++}_{1,n}\quad\text{and}\quad \mathcal{S}^{--+}_{n,1}=\mathcal{S}^{+--}_{1,n}~.
\eeq
Using these expressions we again get a cancelation of all the sub-leading color terms of order $ \mathcal{O}(-\frac{1}{N}) $ and we see the same structure as in \eqref{eq:TJbarCSL} but with specific coefficients. That is the triple-soft limit can be written as
\beqa
\mathcal{T}(z_1)\ahJ^{a}(z_3)&\sim& \frac{2}{9}\frac{\ahJ^{a}(z_3)}{z_{13}^2}+\frac{2}{9}\frac{\bar{z}_{13}\bar{\partial}\ahJ^{a}(z_3)}{z_{13}^2}~.
\label{eq:simultaneousTJbar}
\eeqa

\paragraph{A conjugation operator}
Due to the non-trivial OPE between $J^a$ and $\bar{J}^b$ it is interesting
to define the operator 
 \<
 C(z_1)=\frac{1}{4 \pi i h^{\vee}} \oint \frac{dz_2}{z_{21}} J^a(z_1) \bar J^a(z_2)
 \>
 which satisfies the OPE with current $J^a$
\<
{ C}(z_1){{J}}^a(z_3)\sim \Big[ d_1^2 \frac{  \bar{J}^a(z_3) }{z_{13}^2 }
+ d_1^2  \frac{\bar z_{13}  }{z_{13}^2 }\pb \bar{J}^a(z_3)\Big]+\dots~,
\>
where the constants $d_1$ are those appearing in the $J\bar J$-OPE \eqref{eq:JJbar}
and we see that $C$ essentially acts as a charge conjugation operator. 

Just as for the energy-momentum tensor we can analyze the same OPE by examining
the triple soft-limit of two positive helicity gluons and one negative but now $h_{q_2}=-$
rather than $h_{q_3}$. The calculation is identical to that above and the final result
can be written as
\<
{ \mathcal C}(z_1){{\hJ}}^a(z_3)\sim 2\alpha_{123} \frac{ \bar{\hJ}^a(z_3) }{z_{13}^2 }
+ 2\alpha_{123} \frac{\bar z_{13}  }{z_{13}^2 }\pb \bar{\hJ}^a(z_3)+\dots
\>
where we again see that with the appropriate choice of parameter 
in the multi-soft limit that we find the expression calculated directly from 
the $J\bar J$ OPE. 

\subsubsection*{Acknowledgments}
We would like to thank Einan Gardi for helpful comments  and in particular Jan Plekfa for many useful discussions
 and suggestions. DN would like to thank Congkao Wen for clarifications regarding results in his paper.
The work of TMcL was supported in part by Marie Curie Grant CIG-333851.
DN's research is supported by the SFB 647 ``Raum-Zeit-Materie. Ana\-ly\-tische
und Geometrische Strukturen'' grant.

\bibliographystyle{nb}
\bibliography{soft}

\end{document}